%% file: main.tex
\documentclass[oneside,letter,12pt,pdflatex]{article}
\usepackage[utf8]{inputenc}
\usepackage[margin=1in]{geometry}

\usepackage{float,etoolbox,breakcites,natbib,cite,color,enumerate,amsmath,tabularx,
	dsfont,lscape,tocloft,booktabs,enumitem, graphicx, rotating, pdflscape, comment, threeparttable, adjustbox,nameref, changepage, multirow, bigdelim}
\usepackage{longtable}
\usepackage{afterpage}
\definecolor{winered}{rgb}{0.5,0,0}

\usepackage[colorlinks=true, linkcolor=winered, citecolor=winered, urlcolor=black, bookmarks=true]{hyperref} 
\usepackage{setspace}
\usepackage[bottom]{footmisc}
\usepackage{indentfirst}
\usepackage{caption}
\floatstyle{plaintop}
\restylefloat{table}
\usepackage{subcaption}
\newenvironment{fignote}{\begin{quote}\footnotesize}{\end{quote}}

\title{Instant Loans Can Lift Subjective Well-Being:\\
A Randomized Evaluation of Digital Credit in Nigeria\thanks{
We thank Johannes Haushofer, Jonathan Robinson and Tavneet Suri for helpful conversations and feedback on this study. We are grateful for funding from the Bill and Melinda Gates Foundation, the Center for Effective Global Action, and Innovations for Poverty Action. This study was pre-registered with the AEA RCT Registry (AEARCTR-0005029), and approved by the U.C. Berkeley Committee for the Protection of Human Subjects.}}

\author{
		Daniel Bj{\"o}rkegren\footnote{\small dan@bjorkegren.com} \\ 
			\footnotesize Brown University \ 
		\and
		Joshua E. Blumenstock\footnote{\small jblumenstock@berkeley.edu} \\ 
			\footnotesize U.C. Berkeley \ 
		\and
		Omowunmi Folajimi-Senjobi\footnote{\small iyandaomowunmi@yahoo.com} \\ 
			\footnotesize University of Ibadan \ 
		\and
		Jacqueline Mauro\footnote{\small jacqueline.mauro@berkeley.edu} \\ 
			\footnotesize U.C. Berkeley \ 
		\and
		Suraj R. Nair\footnote{\small suraj.nair@berkeley.edu} \\ 
			\footnotesize U.C. Berkeley \ 
	}

\date{\today}

\begin{document}

\begin{spacing}{1}

\maketitle

\begin{abstract}
Digital loans have exploded in popularity across low- and middle-income countries, providing short term, high interest credit via mobile phones. This paper reports the results of a randomized evaluation of a digital loan product in Nigeria. Being randomly approved for digital credit (irrespective of credit score) substantially increases subjective well-being after an average of three months. For those who are approved, being randomly offered larger loans has an insignificant effect. Neither treatment significantly impacts other measures of welfare. We rule out large short-term impacts -- either positive or negative -- on income and expenditures, resilience, and women’s economic empowerment.




\end{abstract}

\vspace{10mm}

\vspace{3mm}

\vspace{2cm}

\end{spacing}

\clearpage

\input{1.Introduction}
\input{2.Setting}


\input{3.ExpDesign}

\input{4.Results}


\input{5.Conclusion}

\clearpage

\begin{spacing}{1}
\bibliographystyle{aer}
\bibliography{Nigeria}
\end{spacing}

\clearpage

\input{5A.TablesAndFigures}

\appendix

\renewcommand{\thetable}{A\arabic{table}}
\renewcommand{\thefigure}{A\arabic{figure}}
\renewcommand{\thesection}{A\arabic{section}}
\setcounter{table}{0}
\setcounter{figure}{0}
\setcounter{section}{0}



\input{6.Appendix}

\clearpage

\end{document}

%% file: 1.Introduction.tex
\section{Introduction}

Over the last several years, ``digital loans'' have transformed the consumer credit landscape in developing countries. These products, which allow individuals with no formal financial history to access small loans via a  mobile phone, have become enormously popular. In Kenya, a 2018 survey indicated that 27\% of all adults had an outstanding digital credit loan --- much higher than the number who had microfinance loans ($<5\%$) \citep{totolo_kenyas_2018}. In Nigeria, despite the low penetration of formal financial services, over 50 different companies currently offer digital loan products.

In principle, increased access to credit could have positive effects for both households and small enterprises. Despite the strong demand for these loans, critics argue that they may not improve borrowing well-being, since loan terms are opaque and may induce borrowers to fall deep into debt \citep{boston_review}. Interest rates are high -- typically from 138\% to over 1000\% APR \citep{francis_digital_2017} -- and are accompanied by high rates of default  \citep{johnen_promises_2021}. Many have criticised providers for using predatory practices on people with little experience with formal financial products \citep{hindenburg_research_opera_2020}. 
The debate around digital credit in developing countries echoes that surrounding payday lending in wealthy nations, but with higher stakes: these loans are in many cases the only source of formal credit available to billions of people, many of whom live near subsistence levels with little social safety net to fall back on.

This paper presents the results of the first randomized controlled trial to assess the welfare impacts of digital loans. In partnership with a large Financial Services Provider (FSP) in Nigeria, we increased the availability of credit to a random subset of new loan applicants. Some loan applicants who would normally have been denied credit were approved; some loan applicants were randomly offered larger initial loans than they would have otherwise received. After roughly three months, we surveyed 1,618 individuals by phone to study the welfare impact of increased access to digital credit.

Following a pre-registered pre-analysis plan, our analysis produces several results. First, as expected, being auto-approved for a digital loan increased use of formal credit, as measured several months after the initial loan application. Borrowing from the FSP increases by \$30 USD (\$86 PPP) on average. We observe modest substitution away from informal sources of credit, and a statistically insignificant increase in financial health, measured using a standardized 14-question financial health index. For our second treatment, for each dollar increase in the value of the initial loan offer, borrowing from the FSP increases by a total of \$1.24 (\$3.50 PPP)\footnote{Our conversions use the November 2020 exchange rate of \$1 USD = NGN 378.78. When comparing to results in other settings we use the exchange rate \$1 USD PPP = 135.39 NGN.} across all loans. 

Second, being auto-approved for digital credit substantially increases subjective well-being, by 0.281 standard deviations. This effect is large, even in comparison to the effect of cash transfers and multifaceted antipoverty programs, which are 10-20 times more costly to implement \citep{Ridley}.
Most of the improvement comes from reduced depression, as measured by a standard survey module (Patient Health Questionnaire or PHQ-9); it is also supported by a statistically insignificant increase in reported life satisfaction. In contrast to the large treatment effects of auto-approval, offering larger loans has only small and statistically insignificant effects on subjective well-being.

Third, we are able to rule out large effects --- either positive or negative --- on the other key dimensions of welfare that we pre-specified, including income and expenditures, resilience to shocks, and women's economic empowerment. The absence of significant positive effects may not be surprising given the small size of the initial loan offer (these ranged from roughly \$3 to \$35); however, the absence of significant negative effects suggests that the widespread concern over the predatory nature of these loans may be not be justified, at least in our context.

In our final set of results, which was not pre-specified, we explore why access to these small, on-demand loans has such a large impact on subjective well-being.  Respondents report taking out loans in order to meet short run needs. Although our sample size limits our ability to study heterogeneity, we find suggestive evidence that impacts are greatest for individuals who are unemployed or have low credit scores. While the effects are not significant at traditional levels, when an unemployed and employed applicant are both auto-approved, the magnitude of the differential effect is predicted to be large enough to close the gap in subjective well-being between them. It likewise closes the gap associated with low credit scores. These large effects of providing small amounts of liquidity on demand are consistent with a growing literature that suggests that being unable to access small but critical resources in times of need can be damaging for mental health \citep{haushofer_psychology_2014,banerjee_effects_2020}.

The quantitative results from our RCT are also consistent with the qualitative stated opinions of the company's customers: 85\% of our sample reported that loan terms were fair, and 94\% report not regretting taking out a loan from the FSP. Likewise, we do not find evidence of some of the behavioral mistakes that are seen with payday lending. In contrast with \citet{allcott_are_2021}, who find that inexperienced payday lending borrowers in the United States underestimate future borrowing, we find that new applicants actually \textit{over}-estimate future borrowing: applicants predict they have a 62\% chance of borrowing from the partner FSP in the next 30 days on average, but in fact only 42\% do. 

To summarize: we do not find substantial negative effects on borrowers. The few significant effects we observe are positive, and access to digital credit has a substantial positive effect on subjective well-being. One caveat to this generally positive assessment is that our study focuses on the relatively short-term effects of small loans to new borrowers; we cannot say whether different effects would be observed over longer time horizons to long-term customers.




\subsection{Related literature}

This paper complements two recent quasi-experimental evaluations of the welfare impacts of digital credit that exploit discontinuities in loan approvals based on credit score. \citet{bharadwaj_fintech_2019} finds small but generally positive longer-term effects of digital loans in Kenya, particularly with respect to household resilience to shocks. \citet{malawi_kutchova} finds some evidence of positive effects on (self-reported) financial well-being from digital loans in Malawi. They also find that giving borrowers additional information about the (high) fees and risks of default \emph{increased} demand for digital credit.

Our results also relate to a larger literature on the welfare impacts of expanding credit access in low- and middle-income countries. Most relevant to our results, \citet{angelucci_microcredit_2015} and \citet{fernald_small_2008} find that access to microfinance reduces depression, though \citet{fernald_small_2008} also observe it increases stress. We compare our results on subjective well-being to these and other studies in  Section~\ref{sec:discussion}, after presenting our main results. More broadly, empirical studies of credit have highlighted the high returns to capital for small enterprises  \citep{demel,demel2009,karlan2014}, and heterogeneous impacts on household consumption and welfare \citep{banerjee_miracle_2015, tarozzi_impacts_2015, attanasio_impacts_2015, crepon_estimating_2015,karlan_zinman_2010,augsburg_impacts_2015,meager_understanding_2019}. However, digital loans like we study are different from typical microfinance loans: they are much smaller, can be accessed instantaneously, are shorter-term, and typically charge substantially higher interest rates.   

The debate around digital credit also parallels concerns around payday lending in wealthy nations, which also offer repeat, short-term, high-interest rate loans \citep[cf.][]{bhutta_payday_2015}.
That literature documents both positive and negative effects on borrowers \citep{zinman_restricting_2010, melzer_real_2011, melzer_spillovers_2018, morse_payday_2011, morgan_how_2012,  carrell_harms_2014,  bhutta_payday_2015, bhutta_consumer_2016, gathergood_how_2019, skiba_payday_2019}.

%% file: 2.Setting.tex
\section{Setting\label{sec:Setting}}


Our study population is a random sample of new customers on a popular digital credit platform in Nigeria. Nigeria has relatively high rates of financial inclusion relative to neighboring countries: 51\% of adults report using formal financial services \citep{efina_2020}. 
An estimated 89\% of Nigerians own a mobile phone and 28\% of adults report using digital financial services \citep{efina_2020}.

\subsection{The digital credit product}

 Our study examines the welfare impacts of small loans offered by a private financial service provider (FSP) in Nigeria. Consumers can apply for loans via a smartphone application, and the FSP assesses creditworthiness using behavioral data derived from their smartphone \citep[as in][]{bjorkegren_behavior_2020}. Approved applicants are presented with a menu of three loan offers of different value. Applicants must have a bank account to register, but do not need a formal financial history.

 
 In general, loans range from 1,000 Nigerian Naira (NGN), or roughly USD \$2.60, to 200,000 NGN (USD \$528).%
 \footnote{For context, the legally mandated monthly minimum wage in Nigeria was 30,000 NGN.}
Loans are typically due after 28 days, and the interest rates we observe range from 15\% – 22\% per month (implying an annual percentage rate of 195\% to 287\%).  If a borrower does not repay on time, the FSP does not charge a late fee, but if a customer defaults, they are not eligible to apply for future loans from the FSP. If a customer repays, they gradually become eligible for larger loans.
 
In our study sample (N=1,618), the average initial loan amount is approximately NGN 5,600 (\$15); over the roughly 3 months between enrolment and survey, average total borrowing is NGN 21,300 (\$56). Appendix Figure~\ref{loanladder} shows how loan values increase as customers repay prior loans. 7\% of the loans taken out within our sample end in default, and altogether 23\% of borrowers default at least once.  

The product we examine is broadly similar to other digital credit products offered across sub-Saharan Africa \citep{francis_digital_2017}. In particular, it is similar to the M-Shwari loan product in Kenya analyzed in \citet{bharadwaj_fintech_2019}, and the Kutchova product in Malawi analyzed in \citet{malawi_kutchova}, though our FSP's loans tend to be slightly larger.\footnote{For M-Shwari and Kutchova, applicants must have a mobile money account for at least 6 months. Monthly interest rates are 7.5\% and 10\%, respectively, and both lenders charge a late fee (7.5\% and 2.5\%).
In \citet{bharadwaj_fintech_2019}, the average loan size (conditional on borrowing) is approximately \$4.80, and customers borrow roughly \$40 over the 18-month study period. In \citet{malawi_kutchova}, the average loan size is roughly \$4.00, and the average total value of all loans taken out over 3 months is roughly \$18 (conditional on borrowing). }

\subsection{Descriptive evidence}


Qualitative surveys suggest that borrowers like the FSP's product, and demand for loans is high. Among the approved applicants we observe in data from the FSP, 85\% take out a loan. Among those surveyed (details on the survey are provided below), 86\% report that the FSP's loan terms are fair and 94\% report not regretting taking out a loan from the FSP.

We also look for evidence of the sort of behavioral trap observed with payday loans in \citet{allcott_are_2021}, who find that payday borrowers frequently underestimate future borrowing. However, we find that our borrowers actually \textit{over}-estimate future borrowing from the FSP (Appendix Figures~\ref{fig:misprediction} and \ref{fig:misprediction_breakdown}): the average applicant predicts they have a 62\% chance of borrowing from the FSP in the next 30 days, whereas in practice only 42\% borrow within that period.
As in \citet{allcott_are_2021}, the magnitude of misprediction decreases with experience (measured by the number of FSP loans taken out prior to survey).


%% file: 3.ExpDesign.tex
\section{Experimental design and estimation strategy\label{sec:Experiment}}


As part of a research collaboration with the partner FSP, a randomly-selected sample of the FSP's applicants were included in a Randomized Controlled Trial (RCT) to measure the impact of digital loans on well-being. This section describes the experimental design, the data we collected, as well as our estimation strategy, all of which were also pre-registered in our Pre-Analysis Plan (AEARCTR-0005029).

\subsection{Experimental design}

\label{sec:rct}

As part of its normal business operations, the partner FSP frequently runs randomized controlled trials (A/B tests). We worked with the FSP to launch a new RCT, which included a randomly selected 8\% of all new applicants who installed the app between August 2019 to February 2020. In total, 46,937 people were included in the RCT. These participants were cross-randomized across two different treatment arms:

\paragraph{Auto-Approval Treatment (Extensive Margin)} Half of all participants (4\% of all new applicants) were automatically approved for credit, regardless of credit score. The other half (`standard approval' group) were approved only if their credit score exceeded a threshold.\footnote{Applicants in both groups could still be denied credit if their application raised fraud detection flags.} 

\paragraph{Initial Loan Value (Intensive Margin)}
Applicants who were approved received a randomly assigned maximum initial loan offer, selected from NGN 1000, 2000, 5000, 10,000, or 13,000 (between about \$2.75 and \$35.75). Customers who repaid their initial loan on time would subsequently be eligible for future loans according to the FSP's standard loan ladder.



\subsubsection{Subject recruitment, surveys, attrition, and weighting}

All 46,937 applicants who installed our partner FSP's smartphone app during the study period were invited via text message to participate in a phone survey.  Those who opted in by responding to the message were then contacted by the research team, and asked for informed consent to participate in a 25-minute phone survey.

Invitations were staggered over time to ensure that we could quickly follow-up with a phone call to the respondent. To ensure that the different treatment arms were balanced across cohorts (defined as the set of applicants who enrolled in a particular week), we sampled approximately 2,000-2,500 consumers to be contacted each week, roughly three months after enrolment. Surveying began in the week of November 11, 2019, and concluded in the week of February 7, 2020.  The survey gathered details on demographics, household composition, financial behavior, subjective well-being and household decision making. Respondents were compensated with an airtime incentive of NGN 3600 upon completion.

Our main analysis focuses on the subset of 1,618 consumer who responded to the text message and successfully completed the phone survey. We omit from our analysis 439 people who we surveyed but were ineligible for loans because they were never assigned a credit score (typically because they never opened the app after installation, or their data did not successfully upload). Appendix Table \ref{tab:LoanDistsurvey} summarizes treatment assignment for individuals in the final sample.






\label{weighting}
We weight each survey respondent by the inverse probability of being included in our sample, to address two concerns. First, attrition is slightly higher in the auto-approval group (see Appendix Figure \ref{exp_design}). Second, some individuals were not assigned to the standard approval group until slightly later in the enrolment process due to an engineering error. This caused us to survey some of the standard approval group closer to enrolment (3-5 days, on average) than the auto-approval group. We construct probability weights for treatment group interacted with cohort indicators, assuming that consumers who enrolled in a treatment arm in a particular cohort have the same joint probability of being in our final sample. This ensures that the mean weighted distribution of enrolment times are equal. 

\subsubsection{Sample characteristics and balance}

Applicants in our sample are mostly male (76\%), around 30 years of age on average, and educated at the secondary school or university level. Respondents are distributed across the various states of Nigeria, with Lagos having the largest share (33\%). A majority of respondents are employed in either their own business (41\%), or in salaried jobs (39\%). For more details, see Appendix Table~\ref{summ_stats_0}.

Characteristics are balanced across treatment arms (Appendix Table~\ref{balance_table_1618}). We test for balance in a number of ways. First, we examine balance between the auto-approval and standard approval group arms for each individual characteristic (column 2).   
Then, we report the F-stat from a joint test of significance, on all fixed characteristics (column 3). Finally, we test whether the initial loan offer amount is independent of each characteristic (column 6). Overall, we find no significant differences between the average characteristics of the auto-approval and standard approval groups, except for the initial amount offered to applicants in Lagos.\footnote{At the 10\% level, we find that applicants in the auto-approval group are less likely to be in Lagos, and to belong to the Igbo tribe.}

\subsection{Estimation strategy\label{sec3}}

We are interested in understanding how use of digital credit affects the welfare of applicants.
Our two randomized treatments $\boldsymbol{Z}_i$ create exogenous variation in credit access and use. We estimate the impact of these treatments on each outcome $Y_i$ using regressions of the form:

\begin{equation}\label{eq:firststage}
    \mbox{$Y_i$} =  \pi_{0} + \boldsymbol\pi_{1} \boldsymbol{Z}_{i} + \boldsymbol\pi_{2} \boldsymbol{X}_{i}+\nu_{week} + \nu_{enumerator} +\varepsilon_{i}
\end{equation}

To reduce sampling variation, we include a vector of controls ($\boldsymbol{X}_i$: respondent gender, education, ethnicity, location, age, household size, head of household), and fixed effects for week of enrollment and enumerator ($\nu_{week}$ and $\nu_{enumerator}$). All regressions in our analysis are weighted to be representative of the population of first-time borrowers on the FSP's platform (as described in Section~\ref{weighting}).

We have two randomized treatments ($\boldsymbol{Z}_i$). First: a dummy variable indicating whether the respondent is assigned to the auto-approval group. Since this treatment primarily affects the eligibility of applicants whose credit score would normally disqualify them from receiving a loan, we interact it with dummy variables indicating if the respondent would otherwise had been rejected due to having a credit score below the threshold at the time of enrolment (Auto-approval$_{i}$*Under-threshold$_{i}$, and Auto-approval$_{i}$*Over-threshold$_{i}$).%
\footnote{Credit scores change over time and individuals may reapply, so some individuals who are initially above the threshold may still be affected by auto-approval.} 
Second, the value of the randomly assigned initial loan offer for approved applicants (Initial-offer$_{i} \in \{1, 2, 5, 10, 13\}$, in NGN 1000). 
To account for baseline differences between those below the threshold, we include the uninteracted control (Under-threshold$_{i}$), which is not randomly assigned.

%% file: 4.Results.tex
\section{Results\label{sec:Results}}

Our main analysis highlights three sets of results. First, we show how our two randomized treatments -- and in particular the extensive margin that auto-approved loans for applicants with low credit scores -- increased borrowing and affected other financial behaviors of applicants. Second, following our pre-analysis plan, we show how increased access to loans affected several pre-specified indices of welfare; while most effects are statistically insignificant, there are large and significant improvements in subjective well-being. Third, we do a deep dive into the subjective well-being results, to better understand where these effects are coming from, and to contextualize them relative to related interventions.


%

\subsection{Impacts on borrowing}

The effects of the two randomized treatments on the financial behaviors of applicants are shown in Table~\ref{table:table1}.  The first two rows indicate the effect of the extensive margin treatment, being auto-approved for a loan. We show the effect separately for people below (row 1) and above (row 2) the minimum credit score threshold.%
\footnote{The auto-approval treatment could, in principle, affect people who were above the credit score threshold at the time of enrolment if later the individual's credit score decreased or the threshold were raised. In practice, such effects are generally small and insignificant (row 2 of Table~\ref{table:table1}).}
The third row indicates the effect of the intensive margin treatment, the randomly assigned initial loan offer.

Broadly, we find that both treatments increase the amount that applicants borrowed from the partner FSP, but that only the extensive margin treatment increases the likelihood that applicants take out any loan. The auto-approval treatment also affects other aspects of financial behavior, but generally only among applicants below the credit score threshold.

In greater detail, the first column of Table~\ref{table:table1}  reports the effects of both treatments on the total borrowed from the FSP, as observed in administrative data from the FSP, in the period between the initial app installation and the date of the phone survey. For applicants under the threshold at the time of enrollment, auto-approval increases borrowing from FSP by 11,657 NGN (\$30, or \$86 PPP). The next row indicates that, for applicants above the threshold, auto-approval increased borrowing by a statistically insignificant 1,227 NGN.  In the third row, we observe that, for each additional 1,000 NGN offered in the initial loan, borrowing from FSP increases by 1,239 NGN. Since the value of the initial loan ranges from 1,000 to 13,000 NGN, the initial offer treatment induces a predicted difference in borrowing as large as 14,868 NGN. For comparison, individuals in the standard approval group borrow a total of 20,036 NGN on average from the FSP.

The remaining columns of Table~\ref{table:table1} indicate the effects on other financial behaviors. 
In Column 2, we observe that auto-approval increases the proportion of applicants under the threshold who take out \textit{any} loan by 37 percentage points. This effect is driven by having a loan from the FSP: column 3 indicates that auto-approval does not significantly affect the proportion of applicants with a non-FSP loan. The value of the randomly assigned initial offer has no effect on whether either category of loans are taken out (columns 2 and 3).

Columns 4 and 5 of Table~\ref{table:table1} indicate that increased access to digital credit causes applicants to substitute from informal credit towards formal credit. For applicants below the credit score threshold, the auto-approval treatment increased an index of formal borrowing by 0.88 standard deviations and decreased an index of informal borrowing by 0.33 standard deviations. Each index is the average of the z-scores of the number and amount of loans reported taken out in the last 3 months from formal sources (digital credit, bank, microfinance, or cooperative) or informal sources (friends and family, moneylenders, or airtime credit). This substitution away from informal credit is driven by a large reduction in borrowing from friends and family and a small reduction in borrowing from moneylenders --- there is no effect of our treatments on the use of other digital lenders, banks, or cooperatives.%
\footnote{See Appendix Figure~\ref{fig:loan_source_rf}. As context, 80\% of our sample reports borrowing from the partner FSP, and a third of our sample reports borrowing from other digital sources. Borrowing from non-digital formal sources is limited; only 6\% of our sample reports borrowing from a bank, and only 2\% (each) report that they borrow from a micro-finance institution, or from a cooperative. Appendix Table~\ref{summ_stats_1} compares self-reported and administrative data about borrowing.}
We see no significant effect of increasing the initial offer on use of informal credit. 

Both auto-approval and offer amounts significantly increase the applicant's ratio of loans taken out to income (both for one month; column 6) -- the closest our data will allow us to get to a debt/income ratio. Auto-approval increases loans taken out by 7.9 percentage points of income on average for applicants under the threshold; likewise, each additional 1,000 NGN in the initial loan increases this ratio by 0.005. Relative to the mean ratio of 9.8\% of income in standard approval group, these are substantial increases. However, in absolute terms, households have limited use of credit (compare, for instance, to the United States, where the average ratio of household debt payments to income is nearly 100\% \citep{ahn_household_2018}). 

Finally, columns 7-9 indicate that neither treatment had significant effects on the applicant's self-reported income, expenditures, or savings.

\subsection{Welfare impacts}

Beyond the direct impacts on borrowing, we evaluate the impact of access to digital credit on several key dimensions of applicant welfare. We focus on four families of outcomes that we pre-registered and pre-specified prior to conducting the endline survey: Financial health, resilience, women's economic empowerment, and subjective well-being.\footnote{Deviations from the pre-analysis plan are described in Appendix~\ref{sec:deviations_pap}.}  For each family, we focus on summary indices that aggregate a number of related variables. We standardize each variable by subtracting the mean and dividing by the standard deviation of the standard approval group. We then construct the summary index as the mean of the z-scores of the component variables.\footnote{Complete details are provided in Appendix~\ref{var_def}.} In the event that a family has more than one summary index of interest, we report p-values that adjust for multiple hypothesis testing (using the Sidak-Holm adjustment). The impact of our two randomized treatments on these four families of outcome indices are presented in Table~\ref{table:table2_main}. 

\paragraph{Financial health} Results in column 1 indicate that neither expansion of digital credit had an effect on an index of the overall financial health of the applicant, as measured by the respondent's answers to 14 standardized questions \citep{cfpb}. Across the 14 individual questions, the two treatments had generally positive but statistically insignificant effects (Appendix Figure~\ref{fig:fin_outcomes}).

\paragraph{Resilience}
Increased access to digital credit did not significantly impact the applicant's self-reported coping with negative shocks. Column 2 of Table~\ref{table:table2_main} shows the effect on the applicant's ability to experience a negative economic shock without forgoing expenditure or adjusting behavior (based on the questions used in \citet{bharadwaj_fintech_2019}).%
\footnote{This index is defined only for respondents who reported experiencing at least one shock in the three months prior to survey (82\% of the total sample). We find no evidence that our randomized treatments affect the shocks a person experiences (Appendix Figure~\ref{fig:shocks_rf}).} 
Column 3 reports an index of the applicant's ability to pay a large amount in an emergency and manage without income.%

The coefficients on both summary indices are very small and close to zero, and the confidence intervals are fairly tight. We do find some suggestive evidence that auto-approval may help applicants manage shocks without selling household assets (Appendix Figure~\ref{fig:resilience}). 

The result in Column 2 differs from the significant increase in resilience documented by \citet{bharadwaj_fintech_2019}, which finds that households individuals just above the credit score threshold are significantly less likely to report foregoing expenses when faced with a shock (coeff: 0.063, SE: 0.030).%
\footnote{Our pre-specified measure of resilience differs slightly from that used by \citet{bharadwaj_fintech_2019}. In results not shown, we construct a measure of resilience exactly following \citet{bharadwaj_fintech_2019} Table 4A. We find no effect of auto-approval on this measure (coef: 0.007, SE 0.067), but the 95\% confidence intervals overlap, so we are unable to reject that effects are the same size.} 

One important difference between the two contexts is that their study population had digital credit accounts for at least 18 months prior to being surveyed, whereas we observe effects after roughly three months.

\paragraph{Women's economic empowerment}
While several recent studies document the potential for financial services to empower women in developing countries \citep[e.g.,][]{suri2016,field19}, we do not find consistent evidence that increased access to digital credit affected women's economic empowerment. Our focal outcome in column 4 of Table~\ref{table:table2_main} is a summary index that aggregates data on female decision-making, purchases and mobility, and beliefs about female financial autonomy. Beliefs were asked of all respondents. Female behavior is asked of respondents who were either married ($N=551$) or had a live-in partner ($N=56$); mobility was asked also of the women who did not fall into those categories. In all cases, we elicited responses about the affected woman in the household: either the respondent herself (if the respondent is a woman), or the respondent's spouse or live-in partner (if the respondent is a man and has a female partner). On this summary index, we observe statistically insignificant effects.\footnote{Effects are also not significant for the constituent components, shown in Appendix Table~\ref{table2_allwee_appendix}.}


Our results are similar to studies which mostly find no or limited impacts of microcredit on women's empowerment, as summarized in Appendix Figure \ref{Figure:lit_review}. The most straightforward comparison is to studies which report summary indices \citep[i.e.,][]{banerjee_miracle_2015, crepon_estimating_2015, karlan_microcredit_2011}: in all cases, we observe that our confidence intervals overlap.

\subsection{Subjective well-being}

Perhaps our most notable finding is that \emph{access} to digital credit increases subjective well-being substantially, by 0.281 standard deviations (first row of Table~\ref{table:table2_main}, column 5). In contrast, the \emph{amount} that a borrower is allowed to access (row labeled `Initial Offer') has a very small and statistically insignificant effect on subjective well-being. We measure subjective well-being with a summary index that places equal weight on self-reported life satisfaction and a standardized measure of depression, i.e., the nine questions from the Patient Health Questionnaire-9 (PHQ-9). As can be seen in Panel A of Figure~\ref{fig:sub_wellbeing_rf}, the positive effect of loan access on subjective well-being is driven by the PHQ-9 score. Applicants allowed to borrow report being less depressed and report feeling less likely to suffer from poor appetite or overeating. We find small effects on other a number of other components of the PHQ-9, though most of these are only significant at the 10\% level after multiple hypothesis testing adjustments.

\subsection{Discussion}
\label{sec:discussion}

The improvements in subjective well-being we find are large, even relative to those of intensive antipoverty interventions (Appendix Figure~\ref{Figure:lit_review}). For instance, the meta-analysis by \citet{Ridley} finds that multifaceted antipoverty programs increase well-being by 0.17 standard deviations, and cash transfer programs on average increase mental health by 0.105 standard deviations;  \citet{angelucci_microcredit_2015} finds that access to microfinance reduces a depression index by 0.045 standard deviations. Relative to these studies, the effect we observe of 0.281 standard deviations are quite large.%
Perhaps most striking is the fact that these other programs involve much larger transfers: in \citet{Ridley}, the average multifaceted antipoverty program cost \$1,707 PPP and the average cash transfer was \$956 PPP; and the average loan value in \citet{angelucci_microcredit_2015} was \$840 PPP.
In our experiment, respondents borrowed an additional \$30 (\$86 PPP) when assigned to auto-approval --- though some still owed money to the FSP at the time of the survey (46\% of the credit that applicants received had yet to be repaid at the time of the survey)
.
We therefore speculate briefly on the mechanisms that might be driving these substantial effects.

For context, we observe that short-run needs are the most common reasons that our sample reports taking a loan (Figure~\ref{loanpurpose}). These needs include everyday use (49\%), business purposes (42\%), medical expenses (37\%), paying house/shop rent (37\%) and emergencies (20\%).
While the loans are small, such uses could reduce the depressive symptoms observed in Figure~\ref{fig:sub_wellbeing_rf}.%
\footnote{More speculatively, since we observe that the auto-approval treatment (but not the initial offer treatment) causes people to borrow less from friends and family (Appendix Figure~\ref{fig:loan_source_rf}), just as the auto-approval treatment (but not the initial offer treatment) reduces depression, it may be that self-reliance contributes to the increase in subjective well-being.}
This may be especially true in the Nigerian context, where rates of depression and mental disorder are quite high.%
\footnote{According to our endline surveys, 47\% of our sample was screened as having mild depression 
and 10\% as having moderate or severe depression
. More broadly, the 2018-19 Nigerian General Household Survey estimates that 20\% of Nigeria heads of households are chronically depressed \citep{perng_depression_2018}. By comparison, only 12.5\% of individuals in the US reported some level of psychological distress \citep{dhingra2011}.}



There is also suggestive evidence that the effects on subjective well-being are larger for certain types of individuals. Table~\ref{table:het_sub_well} disaggregates the well-being effects by several key sources of heterogeneity in our population. Broadly, we observe that people with lower credit scores and those who are unemployed tend to have lower subjective well-being (rows 8-9), and that access to digital credit improves subjective well-being more for these two groups (rows 2-3). In fact, when individuals of both types are auto-approved, this differential effect is sufficient to close the gap in subjective well-being between unemployed and employed, and more than closes the gap by credit score (row 3 vs. row 9, and row 2 vs. row 8). We also find larger impacts on applicants who own a business (row 4), but little difference in impact by gender (row 5). These findings hold also when the effect is allowed to vary by all of these characteristics simultaneously (column 5). However, this heterogeneity analysis should be viewed speculatively, as it was not a part of our pre-registration plan, and the heterogeneous impact estimates are generally not statistically significant.

More broadly, when trying to understand why increasing access to small loans should have comparable effects on subjective well-being as cash transfer programs, it is important to consider when and to whom benefits are provided. Our experiment considers applicants who requested immediate access to small amounts of credit, and compares those who randomly received loans to those who did not. There has been remarkable demand for this form of product across the developing world. 
In contrast, cash transfer programs often allocate broadly, at times determined by the program. 

The comparison to microcredit is more nuanced. Microcredit typically provides larger loans but has an involved application and repayment process so may be less suited to immediate needs. Other evidence suggests that access to microcredit can reduce symptoms of depression; for instance, \citet{fernald_small_2008} find that increased access to microcredit decreased depressive symptoms from 15\% to 2\% for men (but had no significant effect for women). However, it was accompanied by increased stress. \citet{field_repayment_2012} finds that the design of the microcredit loans can contribute to stress: changing the repayment schedule to monthly rather than weekly resulted in borrowers being 51 percent less likely to report feeling `worried, tense, or anxious' about repaying. Part of the improvements in subjective well-being from digital loans may arise from the security of anticipating that one can borrow in the future as needs arise. Borrowers in our sample anticipate future borrowing, and may view digital loans similarly to a line of credit. 

That we find large effects for extensions of credit that are much smaller than other interventions --- and that we find no effect of providing larger loan offers --- suggests that even small relaxations of credit constraints delivered in moments of need may alleviate some mental health burdens.\footnote{This is also consistent with a recent meta-analysis that finds no association between the size of a cash transfer and its impact on mental health, though that compares sizes in different settings \citep{romero_effect_2021}.} 
Overall, these results are consistent with a growing body of evidence supporting the notion that being unable to access small but critical resources in times of need may be quite damaging for mental health \citep{haushofer_psychology_2014,banerjee_effects_2020}. 




%% file: 5.Conclusion.tex
\section{Conclusion}

The dramatic uptake of digital loans across the developing world suggests a pent-up demand for consumer credit. However, the structure of the digital loan market -- which offers new borrowers short-term loans at high interest rates, and results in high rates of default -- has led to widespread public concern over the potential consequences of this financial transformation. 


Our randomized controlled trial finds that increasing \emph{access} to digital loans can improve subjective well-being among applicants, measured after an average of three months. The magnitude of this effect is similar to that of costly anti-poverty interventions, even though the digital loans we study do not consume government or donor resources. This result highlights how even small relaxations of constraints can substantially improve mental health. At the same time, we do not find that offering applicants \emph{larger} loans has any significant effect. We can also rule out large positive -- and negative -- effects of digital credit access on other dimensions of welfare, including income and expenditures, resilience, and women’s economic empowerment.



%% file: 5A.TablesAndFigures.tex
\section*{Tables and Figures}

\begin{landscape}
    \input{RF_Tables/table1}
\end{landscape}

\input{RF_Tables/table2_main}

\input{RF_Tables/Tab_Het_hyp6_summaryindex}

\begin{landscape}

\begin{figure}[ht!]
  	\centering 
  	\caption{Subjective Well-being}
	\label{fig:sub_wellbeing_rf}
	\includegraphics[width=\linewidth]{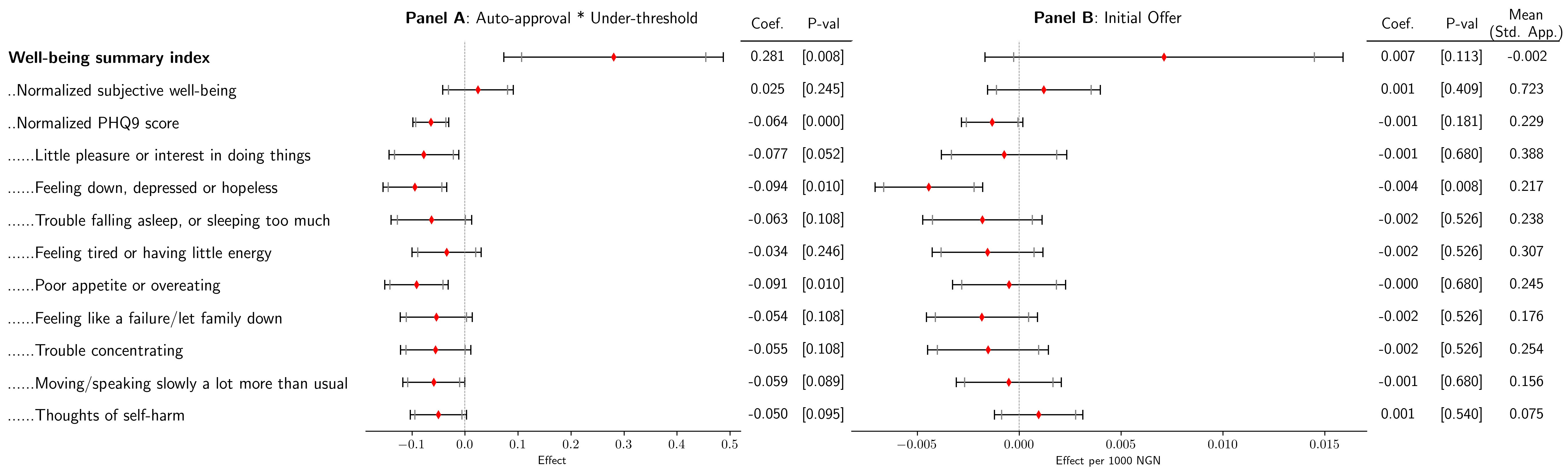}
\begin{fignote}
		\item \emph{\textit{Notes:}} This figure presents reduced form results for measures of subjective well-being. The regression specification is described in Section \ref{sec3}. Black whiskers represent 95\% confidence intervals, and grey whiskers represent 90\% confidence intervals. In each regression, we control for respondent gender, education, ethnicity, location (state), household size, head of household, age, and respondent's credit score status (1=under threshold) at the time of enrolment. We also include enumerator and week of enrolment fixed effects.  Since we have only one main pre-specified outcome (the well-being summary index) for this family, we report the unadjusted p-value for this outcome. We adjust p-values for False Discovery Rate (FDR) for the normalized subjective-wellbeing question, and the normalized PHQ9-score. We also adjust p-values for FDR for the 9 components of the PHQ-9 scale. Note that the PHQ-9 scale can range from 0-27; for ease of visual presentation, we divide the total PHQ-9 score for each respondent by 27, so that the value ranges from 0 to 1. Lower values on the PHQ-9 scale indicate lower levels of depression. Thus negative coefficients for the normalized PHQ-9 score and its components represent improvements in depression.
	\end{fignote}	
\end{figure}

\end{landscape}

\clearpage

%% file: RF_Tables/table1.tex
\begin{table}[H]\renewcommand{\arraystretch}{1.3}\caption{Impacts of Digital Credit Access on Borrowing and Finances}\label{table:table1}\begin{center}\begin{threeparttable}[H]\begin{tabular}{lccccccccc}
\toprule\toprule
{} &                                                                   (1) &                                                   (2) &                                                              (3) &                                                                 (4) &                                                                   (5) &                                                                          (6) &                                                     (7) &                                                      (8) &                                                       (9) \\
& \multicolumn{6}{c}{\centering \small Borrowing}&\multicolumn{3}{c}{\centering \small Finances}\\
\cmidrule(lr){2-7}\cmidrule(lr){8-10}
{} & \multicolumn{1}{m{1.7cm}}{\centering \small Total Borrowing from FSP} & \multicolumn{1}{m{1.5cm}}{\centering \small Any Loan} & \multicolumn{1}{m{1.5cm}}{\centering \small Any Non-FSP Loan} & \multicolumn{1}{m{1.7cm}}{\centering \small Formal Borrowing Index} & \multicolumn{1}{m{1.7cm}}{\centering \small Informal Borrowing Index} & \multicolumn{1}{m{1.5cm}}{\centering \small  Loans Taken Out / Total Income} &     \multicolumn{1}{m{1.5cm}}{\centering \small Income} & \multicolumn{1}{m{1.7cm}}{\centering \small Expenditure} & \multicolumn{1}{m{1.5cm}}{\centering \small Total Saving} \\
{} &                    \multicolumn{1}{m{1.7cm}}{\centering \small (NGN)} &   \multicolumn{1}{m{1.5cm}}{\centering \small (p.p.)} &              \multicolumn{1}{m{1.5cm}}{\centering \small (p.p.)} &                   \multicolumn{1}{m{1.7cm}}{\centering \small (SD)} &                     \multicolumn{1}{m{1.7cm}}{\centering \small (SD)} &                                \multicolumn{1}{m{1.7cm}}{\centering \small } & \multicolumn{1}{m{1.5cm}}{\centering \small (Category)} &     \multicolumn{1}{m{1.7cm}}{\centering \small (Asinh)} &      \multicolumn{1}{m{1.5cm}}{\centering \small (Asinh)} \\
\midrule
\multirow{1}{3cm}{\footnotesize Auto-Approval *Under-threshold}          &                                                               11656.6 &                                                 0.368 &                                                        -0.033 &                                                               0.877 &                                                                -0.326 &                                                                        0.079 &                                                  -0.138 &                                                    0.005 &                                                    -0.754 \\
                                                                         &                                                           (1797.0)*** &                                            (0.051)*** &                                                       (0.061) &                                                          (0.096)*** &                                                            (0.094)*** &                                                                   (0.014)*** &                                                 (0.245) &                                                  (0.254) &                                                   (0.730) \\
\multirow{1}{3cm}{\footnotesize Auto-Approval *Over-threshold}           &                                                                1226.8 &                                                 0.009 &                                                        -0.007 &                                                               0.112 &                                                                -0.040 &                                                                       -0.010 &                                                   0.167 &                                                    0.106 &                                                    -0.426 \\
                                                                         &                                                              (1476.6) &                                               (0.018) &                                                       (0.030) &                                                           (0.051)** &                                                               (0.042) &                                                                      (0.011) &                                                 (0.111) &                                                  (0.106) &                                                   (0.362) \\
\footnotesize Initial Offer                                              &                                                                1239.0 &                                                -0.001 &                                                        -0.003 &                                                              -0.008 &                                                                 0.004 &                                                                        0.005 &                                                   0.008 &                                                   -0.013 &                                                    -0.008 \\
                                                                         &                                                            (136.3)*** &                                               (0.002) &                                                       (0.003) &                                                            (0.005)* &                                                               (0.004) &                                                                   (0.001)*** &                                                 (0.011) &                                                  (0.010) &                                                   (0.034) \\
\midrule
\multirow{1}{3cm}{\footnotesize Mean dep var. (Standard approval group)} &                                                             20036.676 &                                                 0.832 &                                                         0.450 &                                                              -0.005 &                                                                 0.001 &                                                                        0.098 &                                                   2.301 &                                                    8.929 &                                                     5.632 \\
                                                                         &                                                                       &                                                       &                                                               &                                                                     &                                                                       &                                                                              &                                                         &                                                          &                                                           \\
\footnotesize N                                                          &                                                                  1611 &                                                  1611 &                                                          1611 &                                                                1611 &                                                                  1611 &                                                                         1553 &                                                    1553 &                                                     1437 &                                                      1440 \\
\bottomrule\bottomrule
\end{tabular}\footnotesize \begin{tablenotes}\item \textit{Notes:} Each column is a separate regression. Each regression controls for respondent gender, education, head of the household, ethnicity,
location (state), household size, age and respondent's credit score status (1=under threshold) at the time of enrolment. We include enumerator, and week of enrolment fixed effects. 29 respondents did not report their age -- we code these values as 0, 
and include a dummy variable that controls for these missing values. The index variables in columns (4) and (5) include data on the number, and amount of loans from formal and informal sources respectively. In Columns (6) and (7), monthly income is an ordinal variable, 
defined using the following brackets: 0 - 9,999 NGN, 10,000 - 49,999 NGN, 50,000 - 99,999 NGN, 
100,000 - 249,999 NGN and $>$ 250,000 NGN. The outcome variable in Column (6) is the ratio of self-reported borrowing 
(over 3 months, in NGN) and  self-reported income (over 3 months -- we use the midpoint of each respondents monthly 
income brackets, and multiply by 3). Further details on how each outcome variable is constructed are provided in Appendix~\ref{var_def}. The coefficients in Column (7) are from an ordinal logit regression. 
Parentheses contain robust standard errors. 
 \end{tablenotes}\end{threeparttable}\end{center}\end{table}

%% file: RF_Tables/table2_main.tex
\begin{table}[H]\renewcommand{\arraystretch}{1.3}\caption{Impacts of Digital Credit Access on Pre-Specified Measures of Welfare}\label{table:table2_main}\vspace{3mm}\begin{center}\begin{threeparttable}[H]\begin{tabular}{lccccc}
\toprule\toprule
{} &                                                                 (1) &                                                           (2) &                                                                (3) &                                                    (4) &                                                                     (5) \\
&&\multicolumn{2}{c}{\centering \small Resilience}\\
\cmidrule(lr){3-4}
& \small Fin. Health  & \small Resilience & \small Fin. Resilience  & \small WEE  & \small Subj. Well- \\
& \small  Index &\small  Index&\small  Index&\small  Index &\small  Being Index\\
&       &\small  (SD) & \small (SD) &\small  (SD) & \small (SD) \\
\midrule
\multirow{1}{3cm}{\footnotesize Auto-Approval *Under-threshold}          &                                                               0.024 &                                                         0.014 &                                                             -0.069 &                                                 -0.093 &                                                                   0.281 \\
                                                                         &                                                             (0.018) &                                                       (0.080) &                                                            (0.096) &                                                (0.076) &                                                              (0.106)*** \\
                                                                         &                                                             [0.166] &                                                       [0.860] &                                                            [0.718] &                                                [0.226] &                                                                 [0.008] \\
\multirow{1}{3cm}{\footnotesize Auto-Approval *Over-threshold}           &                                                               0.005 &                                                         0.048 &                                                              0.014 &                                                  0.062 &                                                                  -0.013 \\
                                                                         &                                                             (0.008) &                                                       (0.034) &                                                            (0.046) &                                               (0.033)* &                                                                 (0.049) \\
                                                                         &                                                             [0.565] &                                                       [0.286] &                                                            [0.769] &                                                [0.057] &                                                                 [0.796] \\
\footnotesize Initial Offer                                              &                                                               0.001 &                                                        -0.000 &                                                              0.002 &                                                 -0.001 &                                                                   0.007 \\
                                                                         &                                                            (0.001)* &                                                       (0.003) &                                                            (0.004) &                                                (0.003) &                                                                 (0.004) \\
                                                                         &                                                             [0.098] &                                                       [0.947] &                                                            [0.868] &                                                [0.678] &                                                                 [0.113] \\
\midrule
\multirow{1}{3cm}{\footnotesize Mean dep var. \tiny{(Standard approval group)}} &                                                               0.704 &                                                         0.000 &                                                              0.002 &                                                 -0.000 &                                                                  -0.002 \\
\footnotesize $N$ & 1611 &1312 & 1403 & 1611 & 1611 \\
\bottomrule\bottomrule
\end{tabular}\footnotesize \begin{tablenotes}\item \textit{Notes}: Each column is a separate regression. Details on how each index is constructed are provided in Appendix~\ref{var_def}. In brief: (1) includes 14 standardized questions about financial health; (2) includes 7 questions about coping with negative shocks (conditional on having experienced a negative shock); (3) includes two questions about the respondent's ability to access resources in the event of a shock; (4) is an index of Women's Economic Empowerment (WEE) that includes data on female decision-making, purchases and mobility and beliefs about female autonomy; (5) includes a measure of self-reported life satisfaction,
and a standardized measure of depression. 
Each regression controls for respondent gender, education, 
head of the household, ethnicity, location (state), household size, age, and respondent's credit score status (1 = under threshold) at 
the time of enrolment. We include enumerator, and week of enrolment fixed effects. 29 respondents did not report their 
age -- we code these values as 0, and include a dummy variable that controls for these missing values. 
Parentheses contain robust standard errors, and square brackets contain p-values. For resilience outcomes, we report p-values after adjusting for multiple hypothesis testing, using the Sidak-Holm adjustment.  \end{tablenotes}\end{threeparttable}\end{center}\end{table}

%% file: RF_Tables/Tab_Het_hyp6_summaryindex.tex
\begin{table}[H]\renewcommand{\arraystretch}{1.3}\caption{Treatment Effect Heterogeneity: Subjective Well-being}\label{table:het_sub_well}\vspace{3mm}\begin{center}\begin{threeparttable}[H]\begin{tabular}{lccccccccc}
\toprule\toprule
{} &         (1) &         (2) &        (3) &         (4) &        (5) \\
\midrule
\multirow{1}{4.5cm}{\footnotesize Auto-approval * Under-threshold}              &       0.436 &       0.230 &      0.236 &       0.284 &      0.219 \\
                                                                                &  (0.163)*** &    (0.132)* &    (0.152) &     (0.212) &    (0.312) \\
\multirow{1}{5cm}{\footnotesize Auto-approval * Under-threshold * Credit Score} &      -2.316 &             &            &             &     -2.107 \\
                                                                                &     (2.126) &             &            &             &    (2.136) \\
\multirow{1}{5cm}{\footnotesize Auto-approval * Under-threshold * Unemployed}   &             &       0.167 &            &             &      0.259 \\
                                                                                &             &     (0.222) &            &             &    (0.261) \\
\multirow{1}{4.5cm}{\footnotesize Auto-approval * Under-threshold * Business}   &             &             &      0.123 &             &      0.220 \\
                                                                                &             &             &    (0.213) &             &    (0.248) \\
\multirow{1}{4.5cm}{\footnotesize Auto-approval * Under-threshold * Male}       &             &             &            &       0.010 &      0.031 \\
                                                                                &             &             &            &     (0.244) &    (0.241) \\
\multirow{1}{4.5cm}{\footnotesize Auto-approval * Over-threshold}               &      -0.014 &      -0.012 &     -0.014 &      -0.014 &     -0.012 \\
                                                                                &     (0.049) &     (0.049) &    (0.049) &     (0.049) &    (0.049) \\
\footnotesize Initial Offer                                                     &       0.007 &       0.007 &      0.007 &       0.007 &      0.006 \\
                                                                                &     (0.004) &     (0.004) &    (0.004) &     (0.004) &    (0.004) \\
\footnotesize Credit Score                                                      &       0.280 &       0.271 &      0.283 &       0.282 &      0.262 \\
                                                                                &   (0.112)** &   (0.112)** &  (0.112)** &   (0.112)** &  (0.112)** \\
\footnotesize Unemployed                                                        &      -0.129 &      -0.179 &     -0.130 &      -0.128 &     -0.182 \\
                                                                                &    (0.070)* &   (0.081)** &   (0.072)* &    (0.071)* &  (0.081)** \\
\footnotesize Business                                                          &       0.123 &       0.125 &      0.120 &       0.123 &      0.108 \\
                                                                                &  (0.046)*** &  (0.046)*** &  (0.049)** &  (0.046)*** &  (0.049)** \\
\footnotesize Male                                                              &      -0.121 &      -0.120 &     -0.121 &      -0.141 &     -0.143 \\
                                                                                &   (0.060)** &   (0.060)** &  (0.060)** &   (0.066)** &  (0.066)** \\
\midrule
\multirow{1}{3cm}{\footnotesize Mean dep var. (Standard approval group)} &       -0.00 &       -0.00 &      -0.00 &       -0.00 &      -0.00 \\
                                                                         &             &             &            &             &            \\
\footnotesize N                                                          &        1611 &        1611 &       1611 &        1611 &       1611 \\
\bottomrule\bottomrule
\end{tabular}\footnotesize \begin{tablenotes}\item Each column is a separate regression, where we examine heterogeneity in treatment effects. The credit score ranges from 0-1. Business, Male and Unemployed are binary variables. 
Note that in the regressions in columns (1)-(4), 
we also include the under-threshold dummy, and an interaction with the under-threshold dummy and the characteristic of interest.
    In column (5), we include the under-threshold dummy, and interactions of the under-threshold dummy with all heterogeneity characteristics. 
    These coefficients are not displayed in this table.  Parentheses contain robust standard errors. 
 \end{tablenotes}\end{threeparttable}\end{center}\end{table}

%% file: 6.Appendix.tex
\section{Appendix\label{sec:Appendix}}



\begin{figure}[h]
  	\centering 
  	\caption{Loan Ladder}
	\label{loanladder}
	\includegraphics[scale=0.7]{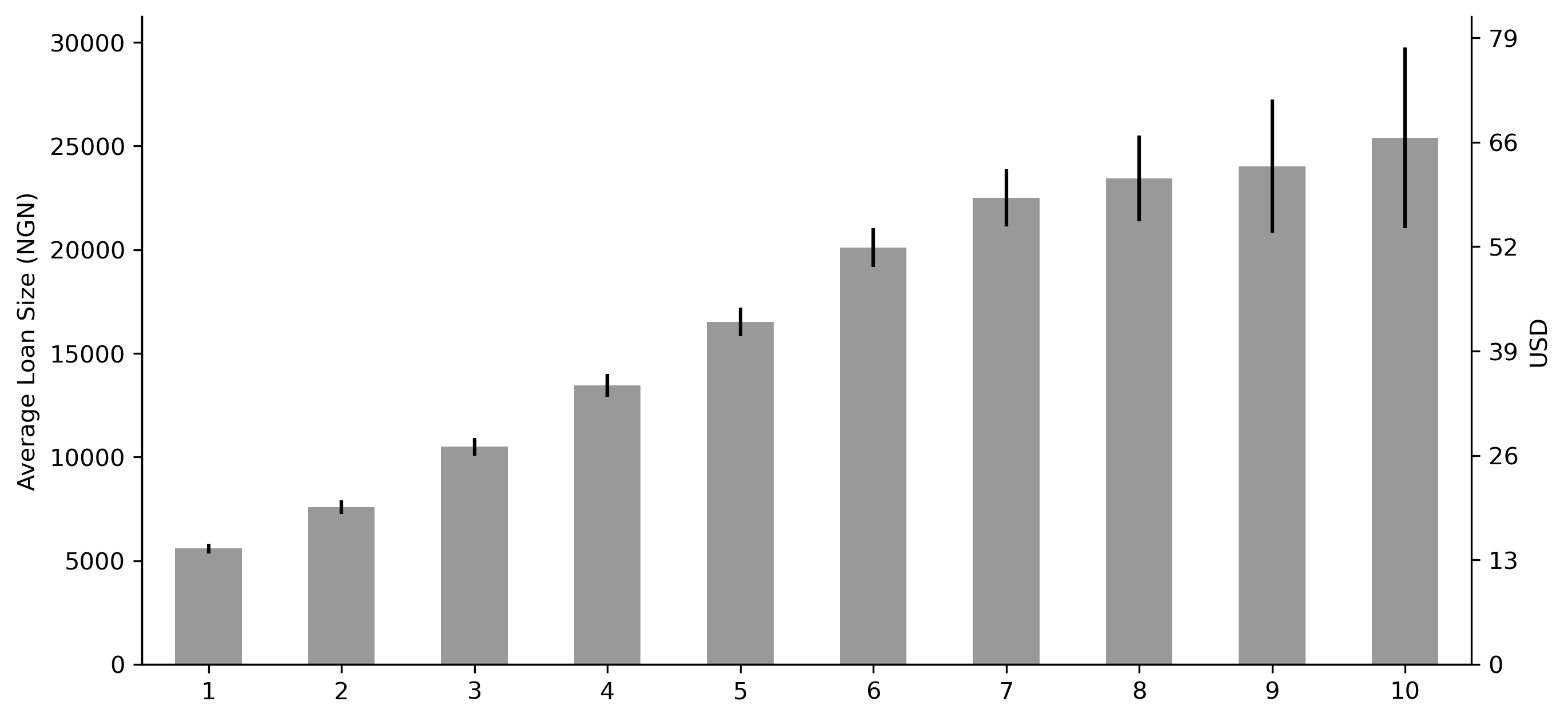} 
	\begin{fignote}
		\item \emph{\textit{Notes:}} This figure presents the mean amount of each loan ($n$th loan, conditional on having borrowed the $(n-1)$th loan), as customers progress up the loan ladder, using loan data provided by the partner FSP for our sample of 1618 customers. For example, the first bar is the mean size of the first loan taken out; roughly 80\% of our sample takes out at least one loan, so the mean is calculated over this set of customers. The second bar is the mean size of the second loan taken out; roughly 70\% of customers take out a second loan, and we use this set of customers to calculate the mean. Black lines indicate 95\% confidence intervals. 
	\end{fignote}	
\end{figure}

\begin{figure}[ht!]
	\centering 
	\caption{Likelihood of Future Borrowing}
	\label{fig:misprediction}
	\includegraphics[scale=0.5]{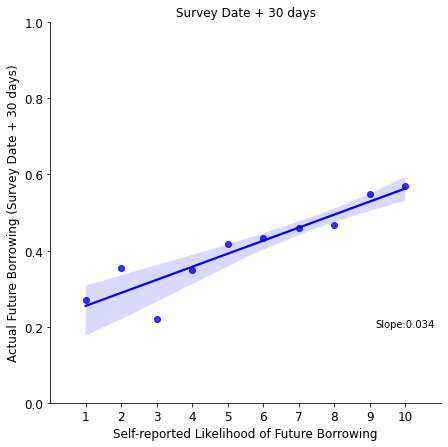}
	\begin{fignote}
		\item \emph{\textit{Notes:}} The outcome variable is based on the following survey question: ``What is the likelihood that you will try to take out another loan from FSP in the next month on a scale from 1 to 10 where 1 is definitely not and 10 is certainly?'' We subtract 1 and divide by 9, so that the value ranges between 0 and 1.
	\end{fignote}
\end{figure}

\begin{figure}[ht!]
  	\centering 
  	\caption{Misprediction of Future Borrowing}
	\label{fig:misprediction_breakdown}
	\includegraphics[scale=0.5]{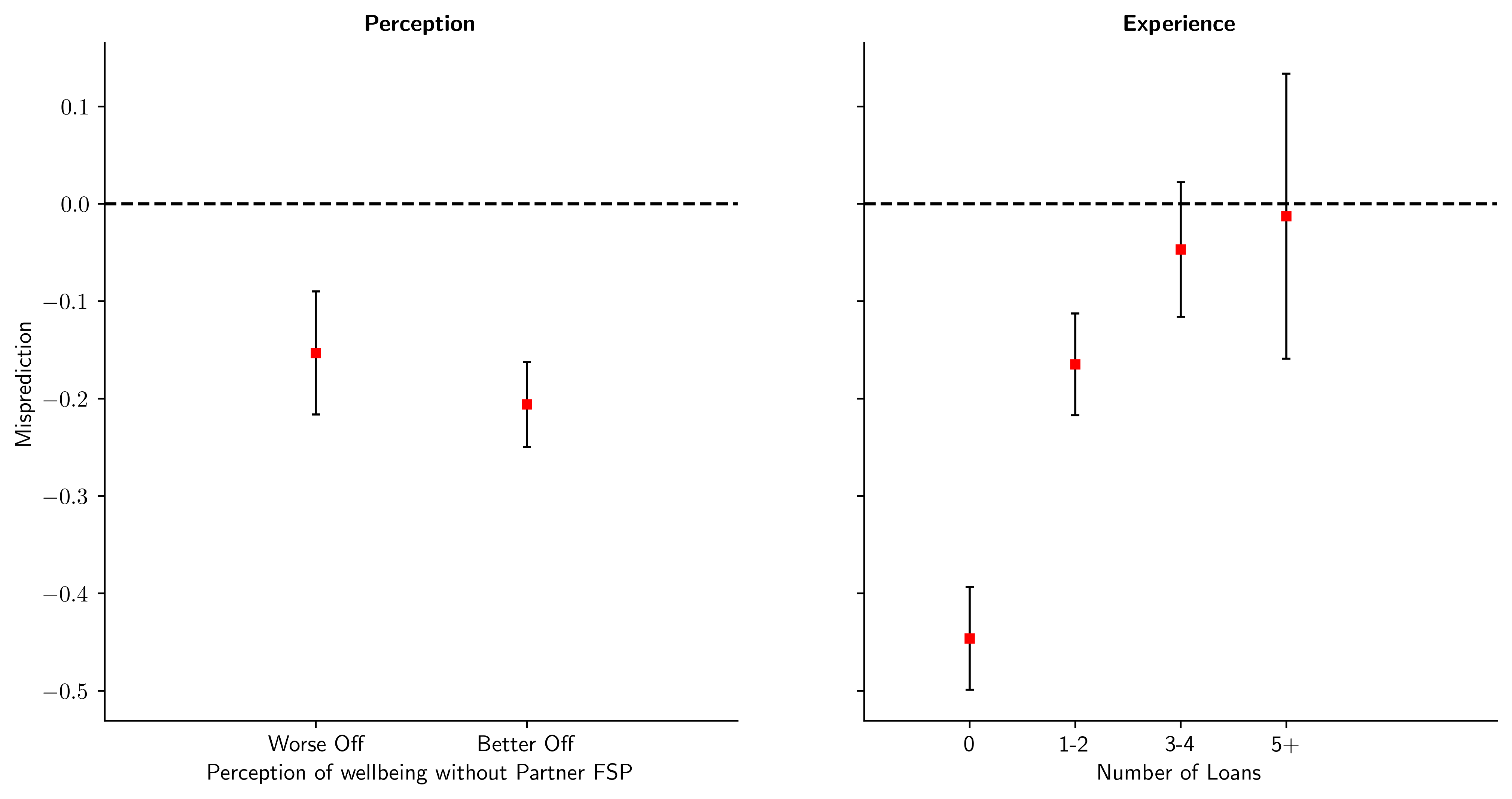}
	\begin{fignote}
		\item \emph{\textit{Notes:}}Misprediction (y-axis) is defined as the actual borrowing probability (based on administrative data 1 month after the survey) minus the predicted likelihood of borrowing in the next month (as defined in Figure~\ref{fig:misprediction}). Perception of wellbeing without partner FSP is self-reported. Number of loans is the total number of loans borrowed from the partner FSP in the 3 months prior to survey. This figure only includes the standard approval group. 
	\end{fignote}	
\end{figure}
\clearpage

\begin{figure}[h]
  	\centering 
  	\caption{Experimental Design}
	\label{exp_design}
	\includegraphics[scale=0.9]{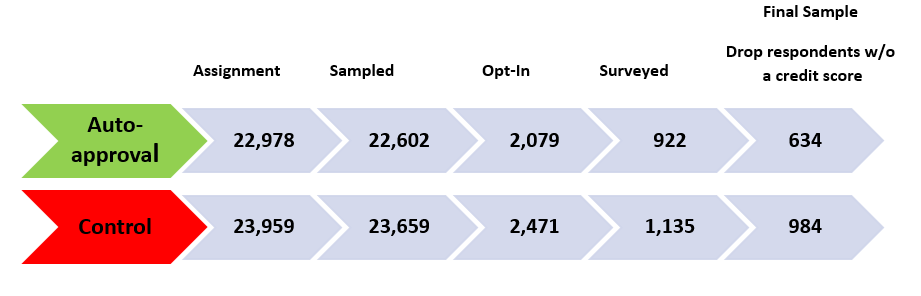}  
\end{figure}

\begin{figure}[ht!]
  	\centering 
  	\caption{Self-reported Loan Sources}
	\label{fig:loan_source_rf}
  \includegraphics[width=\linewidth]{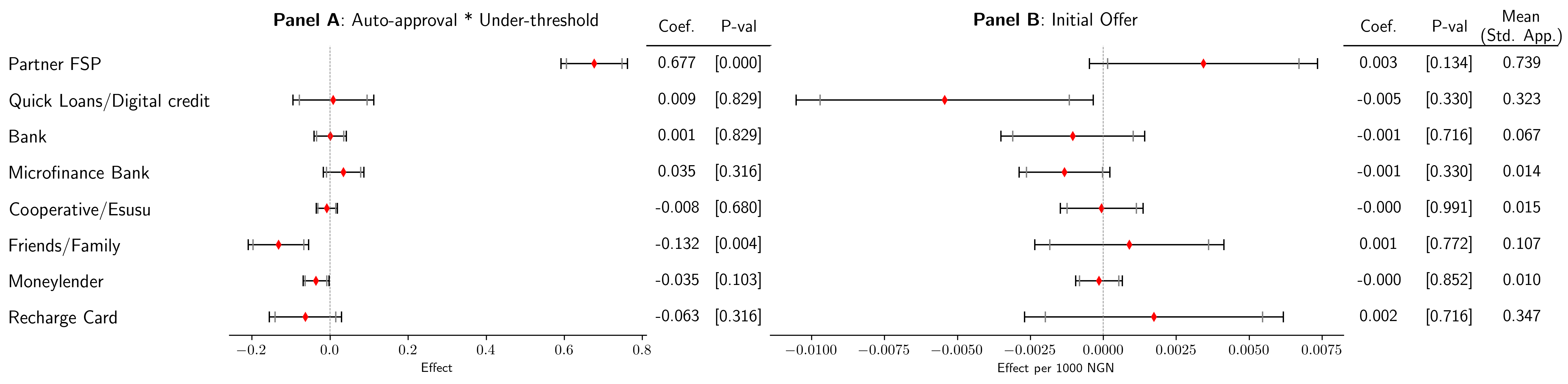}
\begin{fignote}
		\item \emph{\textit{Notes:}} This figure presents reduced form results for self reported borrowing. Each outcome is a dummy variable indicating whether the respondent has borrowed at least once from that source, in the 3 months preceding the survey. The regression specification is described in Section \ref{sec3}. In each regression, we control for respondent gender, education, ethnicity, location (state), household size, head of household, age, and respondent's credit score status (1=under threshold) at the time of enrolment. We also include enumerator and week of enrolment fixed effects. Black whiskers represent 95\% confidence intervals, and grey whiskers represent 90\% confidence intervals.
	\end{fignote}	
\end{figure}

\begin{landscape}
\begin{figure}[ht!]

\caption{Financial Outcomes\label{fig:fin_outcomes}}
\begin{subfigure}{\textwidth}
  \caption{Use of credit, and Savings}
  \includegraphics[width=1.3\linewidth]{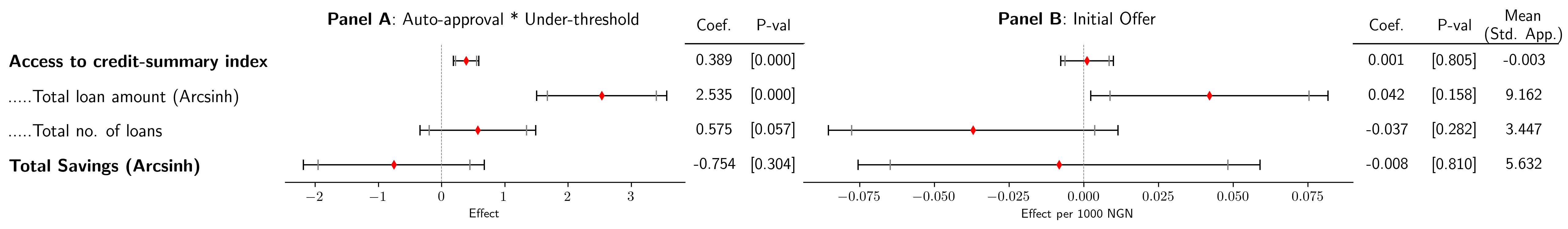}
  \label{fig:sub-first}
\end{subfigure} \\
\begin{subfigure}{\textwidth}
  \caption{Financial Health}
  \includegraphics[width=1.3\linewidth]{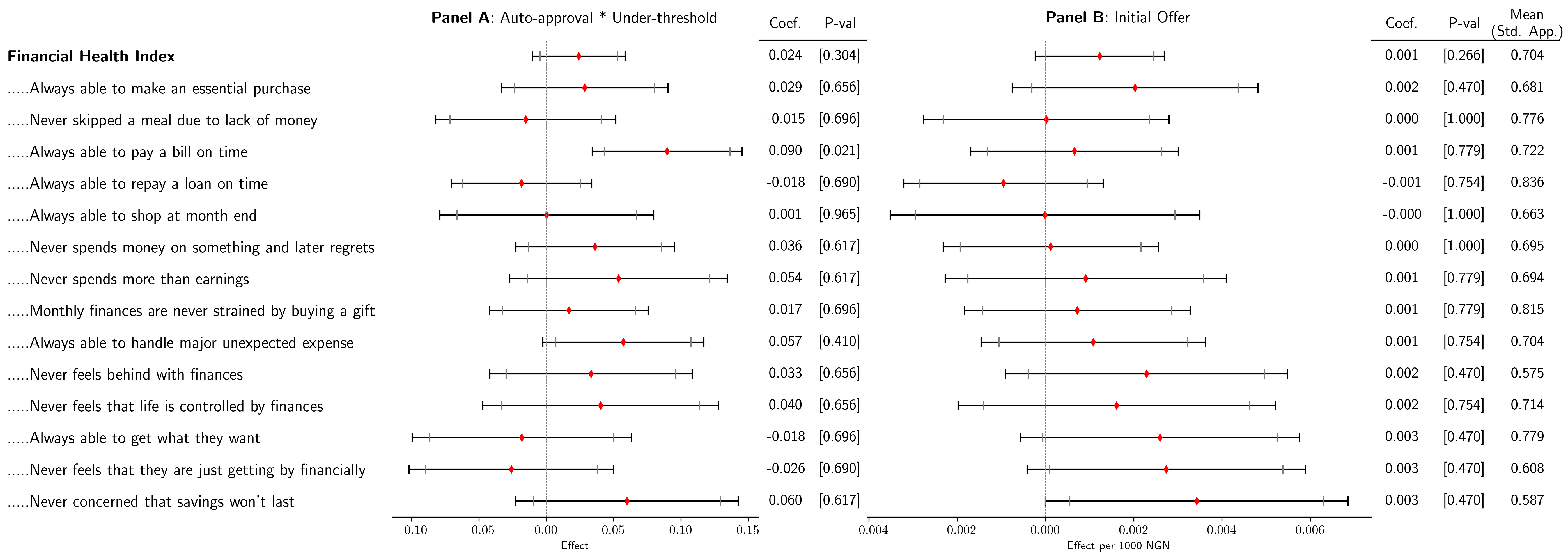}  
  \label{fig:sub-second}
\end{subfigure}
\begin{fignote}
		\item \emph{\textit{Notes:}} This figure presents reduced form results for formal and informal borrowing, and savings outcomes (Row 1), and the financial health score and its sub-components (Row 2). The financial health score aggregates responses from 14 questions that capture various dimensions of the financial health. The financial health score is normalized to range between 0 and 1. The regression specification is described in Section \ref{sec3}. In each regression, we control for respondent gender, education, ethnicity, location (state), household size, head of household, age, and respondent's credit score status (1=under threshold) at the time of enrolment. We also include enumerator and week of enrolment fixed effects. Black whiskers represent 95\% confidence intervals, and grey whiskers represent 90\% confidence intervals. P-values are adjusted for Family Wise Error Rate (FWER) for the main outcomes (bold) and for False Discovery Rate (FDR) for components/ sub-components (indented).
	\end{fignote}
\end{figure}

\end{landscape}

\begin{figure}[ht!]
  	\centering 
  	\caption{Shocks Experienced}
	\label{fig:shocks_rf}
  \includegraphics[width=\linewidth]{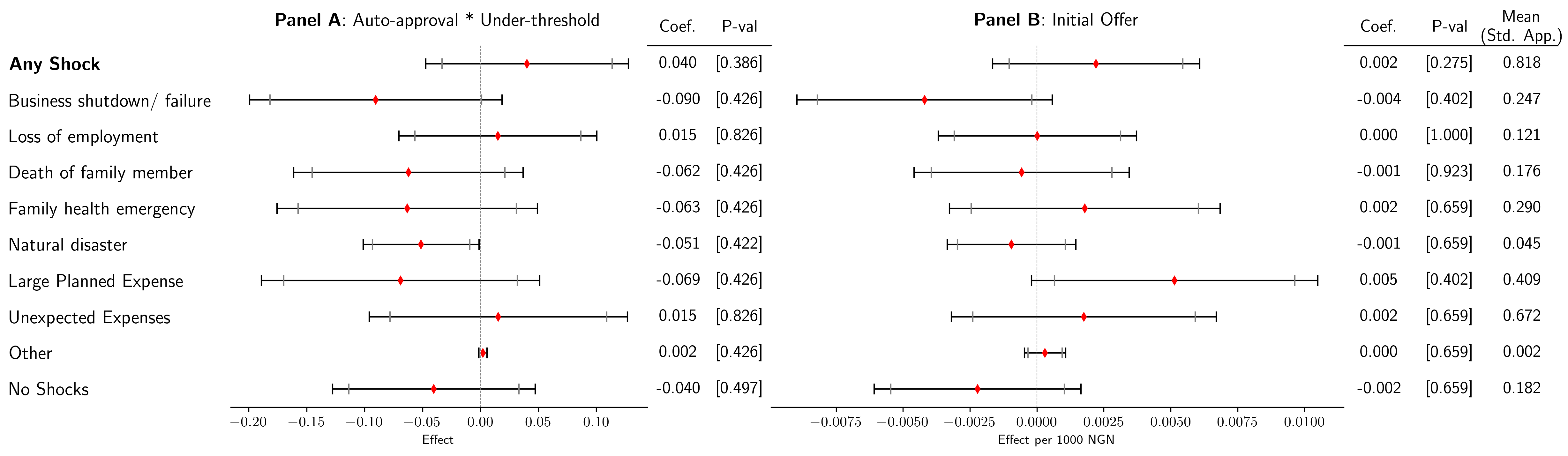}  
  \label{fig:sub-first}
\begin{fignote}
		\item \emph{\textit{Notes:}} This figure presents reduced form results for shocks experienced by the respondent's household in the last 3 months. In bold, we present the coefficients from a dummy variable =1 if the respondent has experienced any of the shocks below it, and 0 otherwise. The regression specification is described in Section \ref{sec3}. In each regression, we control for respondent gender, education, ethnicity, location (state), household size, head of household, age, and respondent's credit score status (1=under threshold) at the time of enrolment. We also include enumerator and week of enrolment fixed effects. Black whiskers represent 95\% confidence intervals, and grey whiskers represent 90\% confidence intervals. All p-values are FDR adjusted. Note that negative coefficients indicate that a respondent is less likely to have experienced a given shock in the last 3 months. 
	\end{fignote}	
\end{figure}

\begin{figure}[ht!]
  	\centering 
  	\caption{Resilience}
	\label{fig:resilience}
	\includegraphics[width=\linewidth]{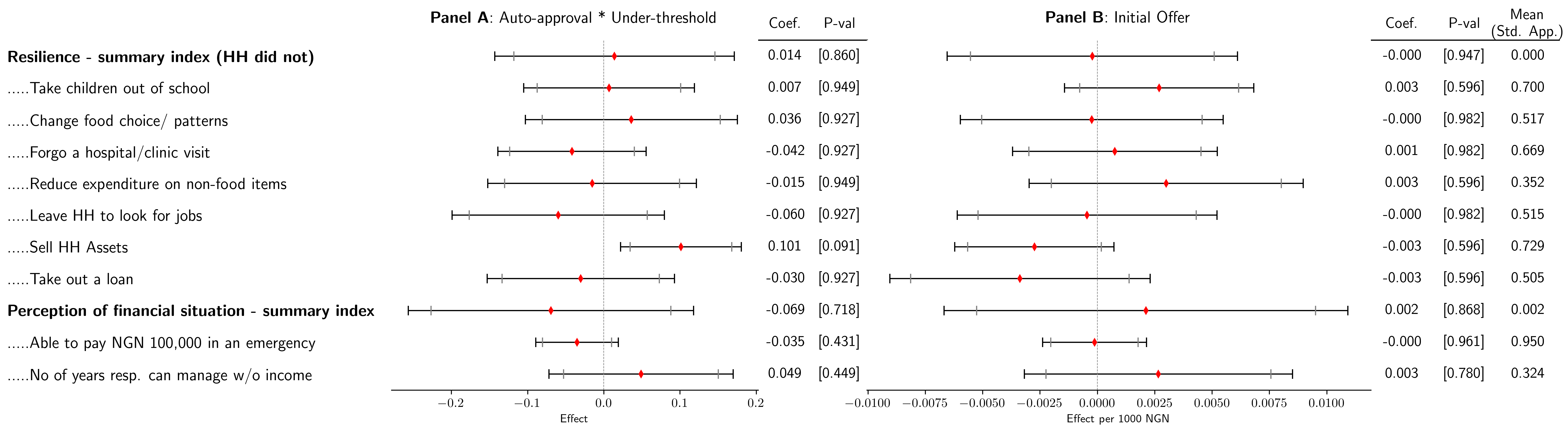}
\begin{fignote}
This figure presents reduced form results for resilience outcomes. The regression specification is described in Section \ref{sec3}. In each regression, we control for respondent gender, education, ethnicity, location (state), household size, head of household, age, and and respondent's credit score status (1=under threshold) at the time of enrolment. We also include enumerator and week of enrolment fixed effects. Black whiskers represent 95\% confidence intervals, and grey whiskers represent 90\% confidence intervals. 
P-values are adjusted for Family Wise Error Rate (FWER) for the main outcomes (bold) and for False Discovery Rate (FDR) for components/ sub-components (indented). 
	\end{fignote}	
\end{figure}

\begin{figure}[h]
\centering
\caption{Effect Size Comparisons: Wellbeing, and Women's Economic Empowerment}
\includegraphics[width=\linewidth]{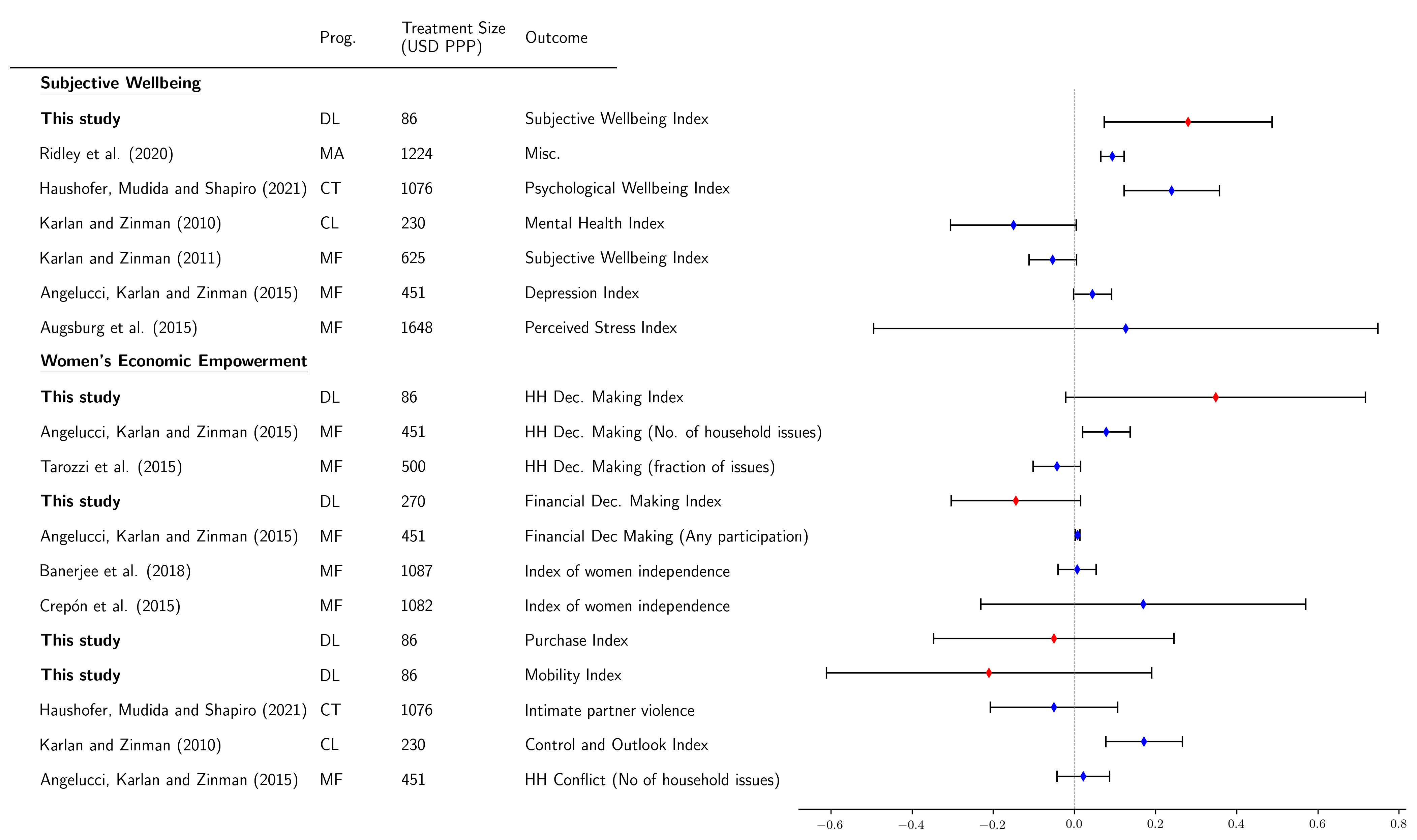}
 \label{Figure:lit_review}
\begin{fignote}\footnotesize \emph{\textbf{Notes: }}This figure plots estimated treatment effects on expenditure from evaluations of digital credit products and various anti-poverty programs. We report coefficients for the auto-approval X under-threshold group from this study in red. Column 2 indicates the type of program: DL refers to Digital Loans, CT refers to Cash Transfers, CL refers to Consumer Loans, MF refers to Microfinance, and
MA refers to Meta-analysis. Column 3 is the size of the treatment in USD PPP. For this study, we report the average borrowing from the partner FSP in the last 3 months. Unless specified otherwise, treatment effects are in standard deviations, and positive coefficients indicate positive outcomes. In \citet{Ridley}, the outcomes considered in their meta-analysis of results include instruments to detect mental illnesses and symptoms of
depression, indices of psychological well-being, and a perceived stress scale. 
\end{fignote}
\end{figure}

\afterpage{
\begin{landscape}
\begin{figure}[ht!]
  	\centering 
  	\caption{Women's Economic Empowerment}
	\label{fig:wee}
\begin{subfigure}{\textwidth}
  \includegraphics[width=1.2\linewidth]{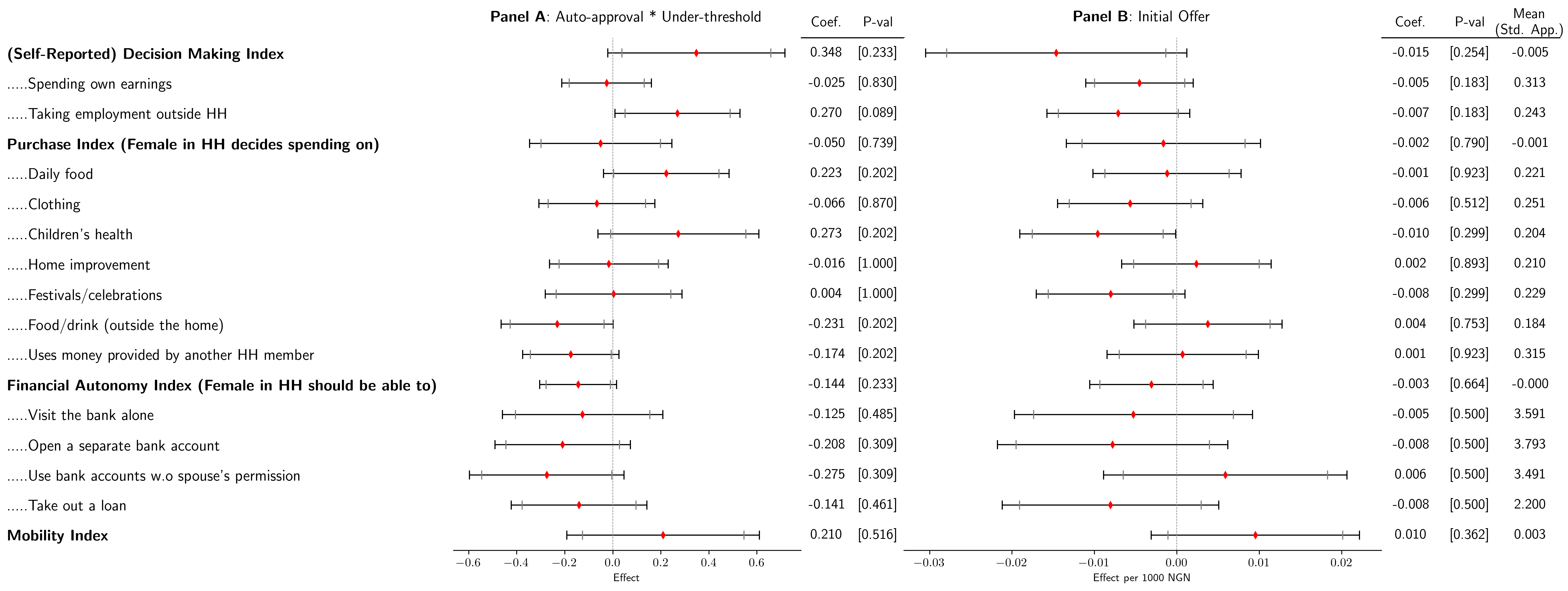}  
  \label{fig:sub-first}
\end{subfigure} \\
\begin{subfigure}{\textwidth}
  \includegraphics[width=1.2\linewidth]{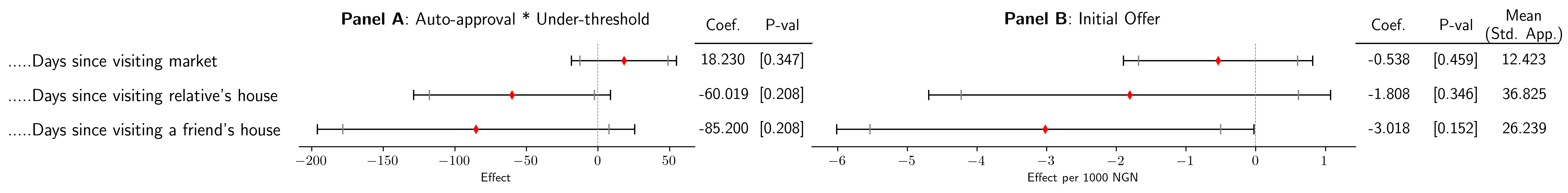}  
  \label{fig:sub-second}
\end{subfigure}
\begin{fignote}
		\item \emph{\textit{Notes:}} This figure presents reduced form results for measures of women's economic empowerment. The regression specification is described in Section \ref{sec3}. In each regression, we control for respondent gender, education, ethnicity, location (state), household size, head of household, age, and respondent's credit score status (1=under threshold) at the time of enrolment. We also include enumerator and week of enrolment fixed effects. Black whiskers represent 95\% confidence intervals, and grey whiskers represent 90\% confidence intervals.
P-values are adjusted for Family Wise Error Rate (FWER) for the main outcomes (bold) and for False Discovery Rate (FDR) for components/ sub-components (indented).
Note that the breakout figure in the second row contains the components of the Mobility Index (presented separately due to the difference in scale), where coefficients are in days and negative values indicate increases in women's mobility.
	\end{fignote}	
\end{figure}
\end{landscape}
}
 \clearpage

\begin{figure}[h]
\centering
\caption{Effect Size Comparisons - Consumption/Expenditure}
\includegraphics[width=\linewidth]{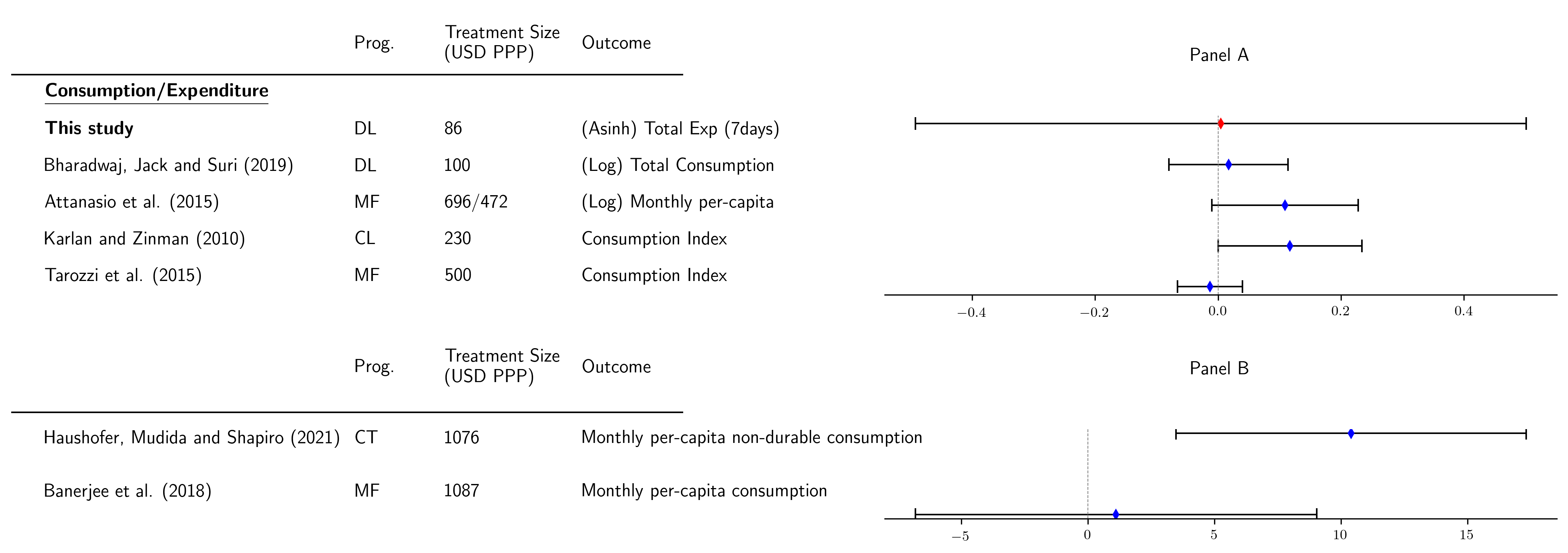}
 \label{Figure:lit_review2}
\begin{fignote}\footnotesize \emph{\textbf{Notes: }}
This figure plots estimated treatment effects on expenditure from evaluations of digital credit products and various anti-poverty programs. We report coefficients for the auto-approval X under-threshold group from this study in red. Column 2 indicates the type of program: DL refers to Digital Loans, CT refers to Cash Transfers, CL refers to Consumer Loans, and MF refers to Microfinance. Column 3 is the size of the treatment in USD PPP. For this study, we report the average borrowing from the partner FSP in the last 3 months. Treatment effects are in standard deviations, and positive coefficients indicate positive outcomes. In Panel B, coefficients are in USD PPP. 
\end{fignote}
\end{figure}

\begin{figure}[h]
  	\centering 
  	\caption{Loan Purpose}
	\label{loanpurpose}
	\includegraphics[scale=0.6]{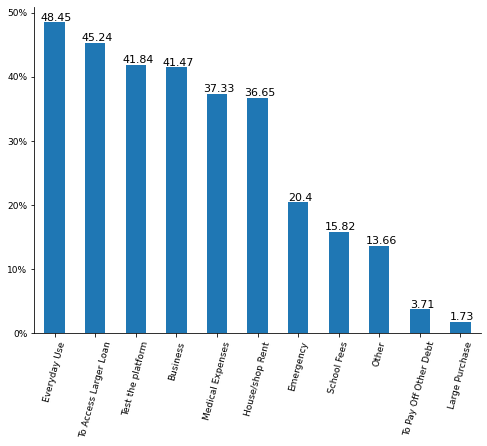}  
	\begin{fignote}
		\item \emph{\textit{Notes:}} Each bar represents the fraction of customers in our sample that report taking out a loan at least once, for that particular purpose, across all loan sources. The exact survey question was: ``For what purpose have you used the money from SOURCE (enumerator, check all that apply, do not read out) [allow multiple selections].''; respondents were asked this question for each type of loan they reported having taken out. 
	\end{fignote}
\end{figure}

\clearpage

\begin{table}[ht!]\renewcommand{\arraystretch}{1.18}
    \centering
        \caption{Treatment Assignment -- Survey Sample}
    
\begin{threeparttable}[H]\begin{tabular}{ccc}\toprule\toprule
\multicolumn{1}{c}{Initial Offer (NGN)}  &    Standard Approval &  Auto-Approval\\ 
\midrule 
1000 &0.18 & 0.17  \\
2000 & 0.20 & 0.16 \\
5000 & 0.20 & 0.22 \\
10000 & 0.21 & 0.21 \\
13000 & 0.22  & 0.24 \\
\midrule
N & 984  & 634 \\
\bottomrule\bottomrule
\end{tabular}\footnotesize \begin{tablenotes}\item Proportions are expressed in terms of the column totals. \end{tablenotes}\end{threeparttable}
    
\label{tab:LoanDistsurvey}
\end{table}

\input{Tables/SummaryStats/summ_stats_0}

\afterpage{
\begin{landscape}
\input{Tables/SummaryStats/Balance_Table1_1618}
\end{landscape}
}

\clearpage
\input{Tables/SummaryStats/summ_stats_1}

\afterpage{
\begin{landscape}
    \input{RF_Tables/table2_allwee_appendix}
\end{landscape}
}

\clearpage

\subsection{Multiple hypothesis testing\label{sec:multipletesting}}

We adjust p-values for multiple hypothesis testing, as described below. (And as specified in our pre-analysis plan). We consider several families of outcomes (e.g., resilience, subjective well-being etc.). For each family, we pre-specified primary outcomes of interest (e.g., the resilience summary index, and financial resilience summary index), which may summarize multiple measures (e.g., multiple questions which measure the applicant's ability to experience a negative economic shock without forgoing expenditure or adjusting behavior). If a family contains only one primary outcome, we do not adjust p-values. If a family contains more than one primary outcome, we report p-values adjusted for Family-wise Error Rate (FWER - the probability that one or more false rejections of the null hypothesis will occur), using the Sidak-Holm adjustment. This adjustment assumes that the tests are not negatively dependent. Under this assumption, for a family of \textit{m} outcomes sorted in ascending order of p-values, the following comparison is sufficient to ensure that the FWER $\leq \alpha$:

\begin{equation}\label{eq:asinh}
    p_m \leq 1-(1-\alpha)^{(1/m)}
\end{equation}

Whenever we present effects on the measures within an outcome (e.g. the individual questions that comprise the resilience summary index in Fig~\ref{fig:resilience}), we adjust p-values for False Discovery Rate (FDR - the expected proportion of the false rejections of the null hypothesis). We use the two stage procedure described in \citet{bky_fdr_twostage} for FDR adjustments. This procedure assumes the independence of p-values (\citet{anderson08} also argues that this procedure works well in the event that p-values are positively dependent). 

\subsection{Deviations from pre-analysis plan}
\label{sec:deviations_pap}
We make the following deviations from our pre-analysis plan:

 \begin{itemize}
    \item Outcomes:
    \begin{itemize}
        \item We report several informative borrowing outcomes in addition to those we pre-specified. We additionally report the extensive margins of taking out any loan and any non-FSP loan, an informal borrowing index created analogously to the formal borrowing index, and the ratio of loans taken out to total income.
        \item We measured income in a categorical variable, because respondents struggled to give exact income amounts during piloting.
        \item The resilience family contains three measures (responses to shocks, raising emergency funds, and meeting basic needs). The pre-analysis plan grouped the first two in one outcome, and the remaining in another outcome. Because the first measure is defined only for respondents who have experienced a shock, we instead leave the first measure as its own outcome (`resilience index') and group the second two measures into an outcome (`financial resilience index').
    \end{itemize}
    \item Specifications:
    \begin{itemize}
        \item We interact the auto-approval treatment with a dummy for whether the applicant is under or above the credit score threshold, which more precisely describes who the treatment is affecting.
        \item The pre-analysis plan suggests instrumental variables (IV) specifications in addition to OLS specifications. We focus on OLS specifications because our two treatments induce different effects. Results with IV specifications are qualitatively similar.
        \item We omit week-of-survey fixed effects; we instead reweight the sample to ensure that the average time between enrolment and survey is the same across auto-approval and standard approval groups.
        \item In addition to the auto-approval and standard approval groups, we also gather survey data from a group of business-as-usual customers. In the pre-analysis plan, we that we would compare both treatment arms against this business-as-usual group in our analysis. However, the business-as-usual group ended up being surveyed roughly 20 days earlier (on average) than the auto-approval and credit-score approval groups. Thus we are unable to make meaningful comparisons between borrowers in the business as usual group, and those in the auto-approval and standard approval arms. As a result of this omission, we do not estimate treatment effects using the specifications outlined in PAP sections 5.3.1 B, and 5.3.2 B. We additionally form z-scores relative to the standard approval group, rather than the business as usual group.
    \end{itemize}
    \item Extensions:
    \begin{itemize}
        \item We attempted to ask about demand for commitment to avoid debt traps, but responses to this question made clear that this question was not understood.
         \item In this paper, we do not present an analysis of heterogeneous treatment effects using machine learning methods. This part of the analysis is still ongoing, and will likely be the subject of a separate paper.
         \item We have not checked the robustness of comparing our index outcomes to ones constructed using Principal Component Analysis (PCA), or using nonparametric methods.
    \end{itemize}

 \end{itemize}




\subsection{Variables \label{var_def}}


Below, we briefly describe the main outcome variables we use for our main analysis. For each outcome, we mention the relevant unit, and the table in which it appears in parentheses.
\\

\textbf{1. Borrowing and Financial outcomes}

\begin{enumerate}[label=\Alph*]
    \item Total borrowing from FSP (NGN, Table 1 column 1): The total amount borrowed from the partner FSP between enrolment and survey.
    \item Any loan (Table 1 column 2): A binary variable equalling one if the respondent self reports taking out a loan from any source, in the last three months. 
    \item Any non-FSP loan (Table 1 column 3): A binary variable equalling one if the respondent self reports taking out a loan from any non-FSP source in the last three months. 
    \item Index of formal borrowing (standard deviations, Table 1 column 4): An equally weighed average of the z-scores of the self-reported i) total number of loans and ii) total amount borrowed, in the last 3 months from formal sources (digital credit, bank, micro-finance, or cooperative). The z-scores are constructed by subtracting the mean of the standard approval group and dividing by the standard deviation of the standard approval group.  
    \item Index of informal borrowing (standard deviations, Table 1 column 5): An equally weighed average of the z-scores of the self-reported i) total number of loans  and ii) total amount borrowed, in the last 3 months from informal sources (friends and family, moneylenders, or airtime credit). The z-scores are constructed by subtracting the mean of the standard approval group and dividing by the standard deviation of the standard approval group.
    \item Ratio of loans taken out to total income (Table 1 column 6): The ratio of self-reported borrowing over the past 3 months in NGN, and self-reported income from the last month multiplied by 3 (we use the midpoint of each individuals income bucket). 
    
    \item Income (categorical, Table 1 column 7): The respondent picks their monthly income bracket from the following categories: $<$N10,000, N10,000-N49,999, N50,000-N99,999, N100,000-N250,000, and $>$N250,000.\footnote{Our pilot studies suggested that eliciting the actual value of monthly income was challenging in our study context, while respondents appeared to be more willing to respond to a categorical question.}
    \item Expenditure (asinh, Table 1 column 8): The inverse hyperbolic sine of the respondent’s total self-reported household expenditure in the last 7 days.
    \item Total Saving (asinh, Table 1 column 9): The inverse hyperbolic sine of the self-reported total amount saved in the three months prior to the survey (Q25). 
    \item Financial Health Index (Table 2 column 1): This index is constructed by aggregating responses from 14 questions that capture various dimensions of the financial health of the respondents. These questions are based on the \citet{cfpb} financial health index, which we piloted and then adjusted in Nigeria prior to our survey. Those responses to each question are collected on a scale of 0-3 (for a maximum of 42); we divide by 42, so that our index ranges between 0 and 1.
\end{enumerate}
  
Following our pre-analysis plan, we group our outcomes into families: subjective well-being, resilience, and women’s economic empowerment. For each family, we specified the primary outcomes in our pre-analysis plan; in most cases, these are index variables following \citet{klk}. 
\\

\textbf{2.	Resilience}

\begin{enumerate}[label=\Alph*]

    \item Resilience index (standard deviations, Table 2, column 2): This index is a standardized equally weighted average of the z-scores of seven questions which capture the respondent’s coping strategies, after having faced a negative shock. The z-scores are constructed by subtracting the mean of the standard approval group and dividing by the standard deviation of the standard approval group. We construct this index for only those respondents who report having faced at least one negative shock in the three months prior to survey. We elicit information on coping strategies using the following question:  In response to adverse events in the last 3 months (ones just named), has your household done any of the following (Yes/No):
    \begin{itemize}
        \item Taken children out of school/ had children sent home from school due to outstanding school fee balance?
        \item Foregone meals, or changed food choice/patterns due to monetary constraints?
        \item Foregone a hospital/clinic visit when a household member was sick, or been unable to pay the full amount needed for some medical treatment?
        \item Reduced expenditure on non-food items?
        \item Had members leave the house to look for jobs?
        \item Sold household assets?
        \item Taken out a loan?
    \end{itemize}
    
In all cases, a `no' response is treated as resilience.

    \item Financial resilience index (standard deviations, Table 2 column 3): This index is a standardized and equally weighted average of the z-scores of two questions, which capture the respondent’s perceptions of their own ability to cope financially with the effects of a hypothetical negative shock. The z-scores are constructed by subtracting the mean of the standard approval group and dividing by the standard deviation of the standard approval group. The two questions used for this index are:
    
    \begin{itemize}
        \item If you had one week to pay 100,000 NGN for an emergency expense, such as a repair or medical bill, who would you turn to, to get the money (read all out, check all that apply)?
        \item  God forbid, if your household stopped getting income from any source, how long could your household easily continue to meet your basic needs for food and housing? (Enter duration in days)
    \end{itemize}

\end{enumerate}

\textbf{3.	Women's Economic Empowerment}
\\

We have four main outcomes that capture various dimensions of women’s economic empowerment. These index variables are based on those used in \citet{field19}, with a few adjustments made to suit our study context. In all cases, the aim is to measure changes in economic empowerment for an adult female in the respondent’s household. If the respondent is male, we thus elicit details about their spouse/ live-in partner. Since 76\% of our sample is male, this is the most common scenario. If the respondent is female, we elicit details about their own perceptions and experiences along the dimensions discussed below. Unless mentioned otherwise below, all questions in this section were administered to only those respondents who report being married or in a live-in relationship.  

\begin{enumerate}[label=\Alph*]
    \item Decision-making index (standard deviations, Appendix Table A5, column 5): This is an equally weighted standardized index of z-scores constructed from two questions which measure the female’s ability to take decisions on how they spend their earnings, and whether they might seek employment outside the household. The z-scores are constructed by subtracting the mean of the standard approval group and dividing by the standard deviation of the standard approval group. 
    The question is: ``Who is responsible for making the following decisions in your household?'' For each option listed below, possible answers include ``you exclusively”, ``mostly you” (coded as one if the respondent is female), ``both you and your spouse/partner evenly” (coded as one for both females and males), ``mostly your spouse/partner”, ``exclusively your spouse/partner” (coded as one if the respondent is male), or ``not applicable”. This variable is missing if the respondent refused to answer, or selectes ``not applicable" 
    \begin{itemize}
        \item how you spend your (worded as ``your spouse/partner spends her'', if the respondent is male) own earnings (meaning income you yourself earn/money you receive (``she earns/money she receives'', if the respondent is male) for benefits)?
        \item whether you take (``your spouse/partner takes'', if the respondent is male) employment outside the household?
    \end{itemize}
    
    \item Purchase Index (standard deviations, Appendix Table A5, column 6): This is an equally weighted standardized index of z-scores constructed from six questions, which measure the female’s ability to make decisions on the purchase of clothing, children’s healthcare, home improvement, festivals, and meals. The z-scores are constructed by subtracting the mean of the standard approval group and dividing by the standard deviation of the standard approval group.
    The main question is: ``Who is responsible for making the following decisions in your household?'' For each option listed below, possible answers include ``you exclusively”, ``mostly you” (coded as one if the respondent is female), ``both you and your spouse/partner evenly” (coded as one for both females and males), ``mostly your spouse/partner”, ``exclusively your spouse/partner” (coded as one if the respondent is male), or ``not applicable”. This variable is missing if the respondent refused to answer, or selects ``not applicable" 
        \begin{itemize}
            \item how much your household spends on clothing,
            \item how much your household spends on your children’s health,
            \item how much your household spends on home improvement,
            \item how much your household spends on festivals and celebrations
            \item how much your household spends on food and drink outside the home

        \end{itemize}
        
    We also include responses from the following question ``When making these purchases do you (``does your spouse/partner'' if the respondent is male) usually use money provided by another household member? (Yes/No)''

    \item Mobility Index (standard deviations, Appendix Table A5, column 7): This is an equally weighted standardized index constructed from the z-scores of three questions, which measure the female’s ability to visit the following locations: a market outside their neighborhood/ village, a relative’s house outside their neighborhood/ village, and a friend’s house for a social visit. For the raw values of each question, lower values indicate a higher frequency of visit/better mobility. The z-scores are constructed by subtracting the mean of the standard approval group and dividing by the standard deviation of the standard approval group. The question is: ``For each location, please tell me approximately how long ago you (your spouse/partner) last visited that location, in days''. The variable is missing if the respondent refused to answer. Note that we administer these questions to all female respondents, irrespective of their relationship status. 
    \begin{itemize}
        \item Market outside neighbourhood/ village
        \item Relative's house outside neighbourhood/ village
        \item A friend's house for a social visit
    \end{itemize}

    \item 	Financial Index (standard deviations, Appendix Table A5, column 8): This is an equally weighted standardized index constructed from the z-scores of four questions, which measure beliefs about whether a female should be able to make the following financial decisions on her own: visiting a bank alone, opening a separate bank account, using their bank account without their partner/husband’s permission, and taking out a loan. Note that we administer these questions to all respondents irrespective of their relationship status. The z-scores are constructed by subtracting the mean of the standard approval group and dividing by the standard deviation of the standard approval group. The questions are: Women should be able to make their own decision to do the following (without needing the permission of their spouse/ live-in partner) [1-Completely disagree, 5-Strongly Agree]
    \begin{itemize}
        \item Visit the bank alone
        \item Open a separate bank account for themselves
        \item Use bank accounts without taking permission from their spouses
        \item Take out a loan
    \end{itemize}
    
\end{enumerate}
We aggregate these four indices to create a single WEE Index (standard deviations, Table 2, column 4): The equally weighted average of the decision-making, purchase, mobility and financial indices. For this variable, we ignore missing values in any of these component indices. 
\\

\textbf{4.	Subjective well-being}

\begin{enumerate}[label=\Alph*]
    \item Index of subjective well-being (standard deviations, Table 2 column 5, Table 3, and Figure 1): The subjective well-being index is a standardized and equally weighted average of two variables: the respondents’ z-score on the PHQ-9 questionnaire, and the z-score of their response to a life satisfaction question, similar to those in the World Values Survey. Note that the respondent’s PHQ-9 score can range from 0-27; for ease of visual presentation, we divide the total PHQ-9 score by 27, so that the value ranges from 0 to 1. A lower PHQ-9 score indicates lower levels of depression. The z-scores are constructed by subtracting the mean of the standard approval group and dividing by the standard deviation of the standard approval group. The life satisfaction question we use is: All things considered, how satisfied are you with your life as a whole these days? (Very happy/ quite happy/ not very happy/ not at all happy)
\end{enumerate}

%% file: Tables/SummaryStats/summ_stats_0.tex
\begin{table}[H]\renewcommand{\arraystretch}{1.3}\caption{Summary Statistics - I}\label{summ_stats_0}\vspace{3mm}\begin{center}\begin{threeparttable}[H]\begin{tabular}{llccccc}
\toprule\toprule
{} & &\multicolumn{2}{c}{(1)} & \multicolumn{2}{c}{(2)} \\
{} & &\multicolumn{2}{c}{Mean} & \multicolumn{2}{c}{Weighted Mean} \\
\midrule
\textsc{\underline{Panel A: Demographics}}      &         &          &               &          \\
\multicolumn{2}{l}{Age}                         &  29.936 &  (8.532) &        29.307 &  (8.384) \\
\multicolumn{2}{l}{Male}                        &   0.758 &  (0.429) &         0.760 &  (0.427) \\
Location:&Lagos                                 &   0.333 &  (0.471) &         0.335 &  (0.472) \\
Education:&Primary                              &   0.007 &  (0.082) &         0.007 &  (0.082) \\
&Secondary                                      &   0.349 &  (0.477) &         0.348 &  (0.476) \\
&HND                                            &   0.093 &  (0.290) &         0.094 &  (0.292) \\
&OND                                            &   0.149 &  (0.356) &         0.149 &  (0.356) \\
&University                                     &   0.357 &  (0.479) &         0.357 &  (0.479) \\
\multicolumn{2}{l}{Head of household}           &   0.447 &  (0.497) &         0.448 &  (0.497) \\
\multicolumn{2}{l}{Household size}              &   5.303 &  (3.199) &         5.246 &  (3.173) \\
Ethnicity:&Yoruba                               &   0.502 &  (0.500) &         0.500 &  (0.500) \\
&Igbo                                           &   0.179 &  (0.383) &         0.179 &  (0.383) \\
&Hausa                                          &   0.043 &  (0.202) &         0.043 &  (0.204) \\
\textsc{\underline{Panel B: Employment/ Misc.}} &         &          &               &          \\
\multicolumn{2}{l}{Primary phone user}          &   0.991 &  (0.093) &         0.992 &  (0.089) \\
\multicolumn{2}{l}{Uses a bank account}         &   0.997 &  (0.056) &         0.997 &  (0.054) \\
Employment:&Self-employed                       &   0.409 &  (0.492) &         0.401 &  (0.490) \\
&Salaried (Full-time)                           &   0.269 &  (0.443) &         0.271 &  (0.445) \\
&Salaried (Part-time)                           &   0.121 &  (0.326) &         0.125 &  (0.331) \\
&Unemployed                                     &   0.201 &  (0.401) &         0.202 &  (0.402) \\
\multicolumn{2}{l}{Days worked last week}       &   3.861 &  (2.418) &         3.871 &  (2.426) \\
\multicolumn{2}{l}{Runs a business}             &   0.551 &  (0.498) &         0.543 &  (0.498) \\
\multicolumn{2}{l}{Aspires to open business}    &   0.806 &  (0.396) &         0.790 &  (0.395) \\
\bottomrule\bottomrule
\end{tabular}\footnotesize \begin{tablenotes}\item Notes: for each variable, column (1) presents the mean and standard deviation. Column (2) presents the weighted mean and standard deviation. The Ordinary National Diploma (OND) is obtained after completing a two-year course at a polytechnic. The Higher National Diploma (HND) requires an additional year of industrial training, or two years of additional coursework. \end{tablenotes}\end{threeparttable}\end{center}\end{table}

%% file: Tables/SummaryStats/Balance_Table1_1618.tex
\begin{table}[H]\renewcommand{\arraystretch}{1.3}\caption{Treatment Arm Balance in Fixed Characteristics}\label{balance_table_1618}\vspace{3mm}\begin{center}\begin{threeparttable}[H]\begin{tabular}{llcccccccc}
\toprule\toprule
{} & &\multicolumn{4}{c}{Auto-approval} & \multicolumn{4}{c}{Initial offer} \\

{} &            &(1) & \multicolumn{2}{c}{(2)} &                                   (3) &           (4) &        (5) &        (6) &   (7) \\
{} & &Mean (Control) & \multicolumn{2}{c}{Difference} &                               F-stat. &     Intercept & \multicolumn{2}{c}{Difference} &     N \\
\midrule
\multicolumn{2}{l}{Male}                         &          0.770 &     -0.025 &   (0.023) &  \hspace{-2em}\rdelim\}{14}{*}[0.9] &         0.779 &     -0.004 &   (0.002)* &  1618 \\
Education:&Primary                               &          0.007 &     -0.001 &   (0.004) &                                     &         0.000 &      0.000 &    (0.000) &  1618 \\
 &Secondary                                      &          0.344 &      0.009 &   (0.025) &                                     &         0.357 &     -0.001 &    (0.003) &  1618 \\
 &HND                                            &          0.103 &     -0.023 &   (0.015) &                                     &         0.086 &      0.001 &    (0.002) &  1618 \\
 &OND                                            &          0.148 &      0.004 &   (0.019) &                                     &         0.156 &      0.001 &    (0.002) &  1618 \\
 &University                                     &          0.356 &      0.002 &   (0.026) &                                     &         0.349 &      0.002 &    (0.003) &  1618 \\
\multicolumn{2}{l}{Household size}               &          5.233 &      0.012 &   (0.165) &                                     &         5.468 &     -0.011 &    (0.018) &  1614 \\
Ethnicity:&Hausa                                 &          0.046 &     -0.007 &   (0.011) &                                     &         0.047 &      0.001 &    (0.001) &  1616 \\
 &Igbo                                           &          0.192 &     -0.034 &  (0.020)* &                                     &         0.193 &     -0.003 &    (0.002) &  1616 \\
 &Yoruba                                         &          0.487 &      0.035 &   (0.027) &                                     &         0.485 &      0.003 &    (0.003) &  1616 \\
\multicolumn{2}{l}{Age}                          &         29.246 &     -0.075 &   (0.444) &                                     &        27.780 &      0.056 &    (0.046) &  1589 \\
\multicolumn{2}{l}{Head of household}            &          0.455 &     -0.020 &   (0.027) &                                     &         0.402 &      0.001 &    (0.003) &  1616 \\
\multicolumn{2}{l}{Married/ has live-in partner} &          0.366 &      0.013 &   (0.026) &                                     &         0.307 &      0.001 &    (0.003) &  1618 \\
Location: & Lagos                                &          0.356 &     -0.049 &  (0.025)* &                                     &         0.261 &      0.006 &  (0.003)** &  1618 \\
\bottomrule\bottomrule
\end{tabular}\footnotesize \begin{tablenotes}\item Note: Column (1) presents the control mean. 
    Column (2) presents coefficients of a ``treatment" dummy, from a WLS regression of each variable on treatment (auto-approval), with no additional controls). Parentheses contain robust standard errors. 
    Column (3) is the F-statistic of a joint test of significance, from a regression of auto-approval on all variables.
    Column (4) is mean of the variable of interest, among those assigned to the lowest initial offer (1000 NGN).
    Column (5) coefficients from a WLS regression of each variable on initial offer (no other controls). Parentheses contain robust standard errors.
    *p$<$.1, **p$<$.05, ***p$<$.01. \end{tablenotes}\end{threeparttable}\end{center}\end{table}

%% file: Tables/SummaryStats/summ_stats_1.tex
\begin{table}[H]\renewcommand{\arraystretch}{1.3}\caption{Summary Statistics - II}\label{summ_stats_1}\vspace{3mm}\begin{center}\begin{threeparttable}[H]\begin{tabular}{llccccc}
\toprule\toprule
{} & \multicolumn{2}{c}{(1)} & \multicolumn{2}{c}{(2)} \\
{} & \multicolumn{2}{c}{Mean} & \multicolumn{2}{c}{Weighted Mean} \\
\midrule
\textsc{\underline{Panel A: Self Reported}}       &            &              &               &              \\
Borrowed from partner FSP                         &      0.797 &      (0.402) &         0.789 &      (0.405) \\
No. of loans                                      &      1.869 &      (1.474) &         1.830 &      (1.467) \\
Total loan amount (NGN)                           &  17377.760 &  (27522.365) &     16419.976 &  (26317.116) \\
Family member borrowed from Partner FSP           &      0.155 &      (0.362) &         0.113 &      (0.312) \\
Made a late repayment                             &      0.342 &      (0.474) &         0.264 &      (0.424) \\
Defaulted                                         &      0.087 &      (0.281) &         0.067 &      (0.248) \\
Will borrow again (1=Most Likely)                 &      0.617 &      (0.332) &         0.616 &      (0.332) \\
Loan terms are fair                               &      0.855 &      (0.352) &         0.829 &      (0.346) \\
Better off without partner FSP                    &      0.681 &      (0.466) &         0.621 &      (0.451) \\
\textsc{\underline{Panel B: Administrative Data}} &            &              &               &              \\
No. of loans                                      &      2.426 &      (1.997) &         2.425 &      (1.999) \\
Total loan amount (NGN)                           &  21284.920 &  (26707.579) &     21202.774 &  (26680.023) \\
No of loan application                            &      3.868 &      (3.445) &         3.859 &      (3.332) \\
No. of rejected applications                      &      1.443 &      (3.332) &         1.434 &      (3.224) \\
Defaulted                                         &      0.229 &      (0.420) &         0.185 &      (0.381) \\
\bottomrule\bottomrule
\end{tabular}\footnotesize \begin{tablenotes}\item Notes: for each variable, column (1) presents the mean and standard deviation. Column (2) presents the weighted mean and standard deviation. All variables in Panel A are based on survey responses. We ask about interactions with the Partner FSP for the three-month period prior to survey. ``Will borrow again'' is based on the following survey question: ``What is the likelihood that you will try to take out another loan from FSP in the next month (30 days) on a scale from 1 to 10 where 1 is definitely not and 10 is certainly?'' We subtract 1 and divide by 9, so that the value ranges between 0 and 1. All variables in Panel B are constructed from administrative data, matched to our sample of survey respondents. We use administrative for the period between enrolment, and survey. \end{tablenotes}\end{threeparttable}\end{center}\end{table}

%% file: RF_Tables/table2_allwee_appendix.tex
\begin{table}[H]\caption{Impacts of Digital Credit Access on Welfare Measures}\label{table2_allwee_appendix}\vspace{3mm}\begin{center}\begin{threeparttable}[H]\begin{tabular}{lccccccccc}
\toprule\toprule
{} &                                                                   (1) &                                                                 (2) &                                                           (3) &                                                                     (4) &                                                                (5) &                                                         (6) &                                                         (7) &                                                          (8) &                                                                     (9) \\
&&&\multicolumn{2}{c}{\centering \small Resilience}&\multicolumn{4}{c}{\centering \small Women's Economic Empowerment}\\
\cmidrule(lr){4-5}\cmidrule(lr){6-9}
{} & \multicolumn{1}{m{1.7cm}}{\centering \small Total Borrowing from FSP} & \multicolumn{1}{m{1.5cm}}{\centering \small Fin. Health Index} & \multicolumn{1}{m{1.5cm}}{\centering \small Resilience Index} & \multicolumn{1}{m{1.7cm}}{\centering \small Fin. Resilience Index} & \multicolumn{1}{m{1.5cm}}{\centering \small Decision Making Index} & \multicolumn{1}{m{1.5cm}}{\centering \small Purchase Index} & \multicolumn{1}{m{1.5cm}}{\centering \small Mobility Index} & \multicolumn{1}{m{1.0cm}}{\centering \small Fin. Index} & \multicolumn{1}{m{1.7cm}}{\centering \small Subj. Well-being Index} \\
{} &                    \multicolumn{1}{c}{\centering \small (NGN)} &                       \multicolumn{1}{c}{\centering \small } &             \multicolumn{1}{c}{\centering \small (SD)} &                       \multicolumn{1}{m{1.5cm}}{\centering \small (SD)} &                  \multicolumn{1}{c}{\centering \small (SD)} &           \multicolumn{1}{m{1.5cm}}{\centering \small (SD)} &           \multicolumn{1}{c}{\centering \small (SD)} &            \multicolumn{1}{c}{\centering \small (SD)} &                       \multicolumn{1}{c}{\centering \small (SD)} \\
\midrule
\multirow{1}{3cm}{\footnotesize Auto-Approval *Under-threshold}          &                                                               11656.6 &                                                               0.024 &                                                         0.014 &                                                                  -0.069 &                                                              0.348 &                                                      -0.050 &                                                       0.210 &                                                       -0.144 &                                                                   0.281 \\
                                                                         &                                                           (1797.0)*** &                                                             (0.018) &                                                       (0.080) &                                                                 (0.096) &                                                           (0.188)* &                                                     (0.151) &                                                     (0.204) &                                                     (0.081)* &                                                              (0.106)*** \\
                                                                         &                                                                 [0.0] &                                                             [0.166] &                                                       [0.860] &                                                                 [0.718] &                                                            [0.233] &                                                     [0.739] &                                                     [0.516] &                                                      [0.233] &                                                                 [0.008] \\
\multirow{1}{3cm}{\footnotesize Auto-Approval *Over-threshold}           &                                                                1226.8 &                                                               0.005 &                                                         0.048 &                                                                   0.014 &                                                             -0.028 &                                                       0.033 &                                                       0.012 &                                                        0.062 &                                                                  -0.013 \\
                                                                         &                                                              (1476.6) &                                                             (0.008) &                                                       (0.034) &                                                                 (0.046) &                                                            (0.080) &                                                     (0.060) &                                                     (0.058) &                                                      (0.039) &                                                                 (0.049) \\
                                                                         &                                                                 [0.4] &                                                             [0.565] &                                                       [0.286] &                                                                 [0.769] &                                                            [0.927] &                                                     [0.923] &                                                     [0.927] &                                                      [0.382] &                                                                 [0.796] \\
\footnotesize Initial Offer                                              &                                                                1239.0 &                                                               0.001 &                                                        -0.000 &                                                                   0.002 &                                                             -0.015 &                                                      -0.002 &                                                       0.010 &                                                       -0.003 &                                                                   0.007 \\
                                                                         &                                                            (136.3)*** &                                                            (0.001)* &                                                       (0.003) &                                                                 (0.004) &                                                           (0.008)* &                                                     (0.006) &                                                     (0.006) &                                                      (0.004) &                                                                 (0.004) \\
                                                                         &                                                                 [0.0] &                                                             [0.098] &                                                       [0.947] &                                                                 [0.868] &                                                            [0.254] &                                                     [0.790] &                                                     [0.362] &                                                      [0.664] &                                                                 [0.113] \\
\midrule
\multirow{1}{3cm}{\footnotesize Mean dep var. \tiny{(Standard approval group)}} &                                                             20036.676 &                                                               0.704 &                                                         0.000 &                                                                   0.002 &                                                             -0.005 &                                                      -0.001 &                                                       0.003 &                                                       -0.000 &                                                                  -0.002 \\
                                                                         &                                                                       &                                                                     &                                                               &                                                                         &                                                                    &                                                             &                                                             &                                                              &                                                                         \\
\footnotesize N                                                          &                                                                  1611 &                                                                1611 &                                                          1312 &                                                                    1403 &                                                                578 &                                                         514 &                                                         515 &                                                         1611 &                                                                    1611 \\
\bottomrule\bottomrule
\end{tabular}\footnotesize \begin{tablenotes}\item Notes: Each column is a separate WLS regression. Details on how each index is constructed are provided in Appendix~\ref{var_def}. In brief: (2) includes 14 standardized questions about financial health; (3) includes 7 questions about coping with negative shocks (conditional on having experienced a negative shock); (4) includes two questions about the respondent's ability to access resources in the event of a shock; (5) - (8) are indices of Women's Economic Empowerment (WEE) that includes data on female decision-making (2 questions), purchases (6 questions) and mobility (3 questions) and beliefs about female autonomy (4 questions); (9) includes a measure of self-reported life satisfaction,
and a standardized measure of depression (9 questions). Each regression controls for respondent gender, education, 
head of the household, ethnicity, location (state), household size, age, and respondent's credit score status (1 = under threshold) at 
the time of enrolment. We include enumerator, and week of enrolment fixed effects. 29 respondents did not report their 
age -- we code these values as 0, and include a dummy variable that controls for these missing values. 
Parentheses contain robust standard errors, and square brackets contain p-values. For resilience and women's economic empowerment outcomes, we report p-values after adjusting for multiple hypothesis testing, using the Sidak-Holm adjustment.  \end{tablenotes}\end{threeparttable}\end{center}\end{table}

%% file: Nigeria.bib
@Preamble{ " \newcommand{\noop}[1]{} " }

@article{dhingra2011,
author = {Dhingra, Satvinder S. and Zack, Matthew M. and Strine, Tara W. and Druss, Benjamin G. and Berry, Joyce T. and Balluz, Lina S.},
title = {Psychological Distress Severity of Adults Reporting Receipt of Treatment for Mental Health Problems in the BRFSS},
journal = {Psychiatric Services},
volume = {62},
number = {4},
pages = {396-403},
year = {2011},
doi = {10.1176/ps.62.4.pss6204\_0396},
    note ={PMID: 21459991},

URL = { 
        https://ps.psychiatryonline.org/doi/abs/10.1176/ps.62.4.pss6204_0396
    
},
eprint = { 
        https://ps.psychiatryonline.org/doi/pdf/10.1176/ps.62.4.pss6204_0396
    
}
}

@article{field_repayment_2012,
	title = {Repayment {Flexibility} {Can} {Reduce} {Financial} {Stress}: {A} {Randomized} {Control} {Trial} with {Microfinance} {Clients} in {India}},
	volume = {7},
	issn = {1932-6203},
	shorttitle = {Repayment {Flexibility} {Can} {Reduce} {Financial} {Stress}},
	url = {https://journals.plos.org/plosone/article?id=10.1371/journal.pone.0045679},
	doi = {10.1371/journal.pone.0045679},
	language = {en},
	number = {9},
	urldate = {2022-02-08},
	journal = {PLOS ONE},
	author = {Field, Erica and Pande, Rohini and Papp, John and Park, Y. Jeanette},
	month = sep,
	year = {2012},
	note = {Publisher: Public Library of Science},
	pages = {e45679},
}

@article{haushofer_psychology_2014,
	title = {On the psychology of poverty},
	copyright = {Copyright © 2014, American Association for the Advancement of Science},
	url = {https://www.science.org/doi/abs/10.1126/science.1232491},
	doi = {10.1126/science.1232491},
	language = {EN},
	urldate = {2022-02-08},
	journal = {Science},
	author = {Haushofer, Johannes and Fehr, Ernst},
	month = may,
	year = {2014},
	note = {Publisher: American Association for the Advancement of Science},
}

@article{fernald_small_2008,
	title = {Small individual loans and mental health: a randomized controlled trial among {South} {African} adults},
	volume = {8},
	issn = {1471-2458},
	shorttitle = {Small individual loans and mental health},
	url = {https://doi.org/10.1186/1471-2458-8-409},
	doi = {10.1186/1471-2458-8-409},
	language = {en},
	number = {1},
	urldate = {2022-02-08},
	journal = {BMC Public Health},
	author = {Fernald, Lia CH and Hamad, Rita and Karlan, Dean and Ozer, Emily J. and Zinman, Jonathan},
	year = {2008},
	pages = {409},
}

@misc{ahn_household_2018,
	title = {Household {Debt}-to-{Income} {Ratios} in the {Enhanced} {Financial} {Accounts}},
	url = {https://www.federalreserve.gov/econres/notes/feds-notes/household-debt-to-income-ratios-in-the-enhanced-financial-accounts-20180109.htm},
	urldate = {2022-02-09},
	journal = {FEDS Notes},
	author = {Ahn, Michael and Batty, Mike and Meisenzahl, Ralf R.},
	month = jan,
	year = {2018},
}

@misc{perng_depression_2018,
	title = {Depression and its links to conflict and welfare in {Nigeria}},
	url = {https://blogs.worldbank.org/nasikiliza/depression-and-its-links-to-conflict-and-welfare-in-nigeria},
	language = {en},
	urldate = {2022-02-09},
	journal = {Nasikiliza},
	author = {Perng, Julia and McGee, Kevin and Oseni, Gbemisola and Sato, Ryoko and Tanaka, Tomomi},
	month = jan,
	year = {2018},
}

@techreport{romero_effect_2021,
	type = {Working {Paper}},
	title = {The {Effect} of {Economic} {Transfers} on {Psychological} {Well}-{Being} and {Mental} {Health}},
	url = {https://haushofer.ne.su.se/publications/Romero_et_al_2021.pdf},
	urldate = {2022-01-30},
	author = {Romero, Jimena and Esopo, Kristina and McGuire, Joel and Haushofer, Johannes},
	year = {2021},
}

@article{banerjee_effects_2020,
	title = {Effects of a {Universal} {Basic} {Income} during the pandemic},
	journal = {UC San Diego Technical Report},
	author = {Banerjee, Abhijit and Faye, Michael and Krueger, Alan and Niehaus, Paul and Suri, Tavneet},
	year = {2020},
}

@article{suri2016,
author = {Tavneet Suri  and William Jack },
title = {The long-run poverty and gender impacts of mobile money},
journal = {Science},
volume = {354},
number = {6317},
pages = {1288-1292},
year = {2016},
doi = {10.1126/science.aah5309}, 
URL = {https://www.science.org/doi/abs/10.1126/science.aah5309},
eprint = {https://www.science.org/doi/pdf/10.1126/science.aah5309}
}

@Article{karlan2014,
  author={Dean Karlan and Robert Osei and Isaac Osei-Akoto and Christopher Udry},
  title={{Agricultural Decisions after Relaxing Credit and Risk Constraints}},
  journal={The Quarterly Journal of Economics},
  year=2014,
  volume={129},
  number={2},
  pages={597-652},
  month={},
  doi={},
  url={https://ideas.repec.org/a/oup/qjecon/v129y2014i2p597-652.html}
}

@article{bky_fdr_twostage,
 ISSN = {00063444},
 URL = {http://www.jstor.org/stable/20441303},
 author = {Benjamini, Yoav  and Krieger , Abba M. and Yekutieli, Daniel},
 journal = {Biometrika},
 number = {3},
 pages = {491--507},
 publisher = {[Oxford University Press, Biometrika Trust]},
 title = {Adaptive Linear Step-up Procedures That Control the False Discovery Rate},
 volume = {93},
 year = {2006}
}

@techreport{efina_2020,
	title = {{EFInA} Access to Financial Services in Nigeria, 2020},
	url = {https://a2f.ng/wp-content/uploads/2021/06/A2F-2020-Final-Report.pdf},
	language = {en},
	urldate = {2021-06-30},
	author = {{EFInA}},
	year = {2021},
}

@article{demel,
 ISSN = {00335533, 15314650},
 URL = {http://www.jstor.org/stable/40506211},
 author = {de Mel, Suresh and McKenzie, David and Woodruff, Christopher},
 journal = {The Quarterly Journal of Economics},
 number = {4},
 pages = {1329--1372},
 publisher = {Oxford University Press},
 title = {Returns to Capital in Microenterprises: Evidence from a Field Experiment},
 volume = {123},
 year = {2008}
}

@article{demel2009,
Author = {de Mel, Suresh and McKenzie, David and Woodruff, Christopher},
Title = {Are Women More Credit Constrained? Experimental Evidence on Gender and Microenterprise Returns},
Journal = {American Economic Journal: Applied Economics},
Volume = {1},
Number = {3},
Year = {2009},
Month = {July},
Pages = {1-32},
DOI = {10.1257/app.1.3.1},
URL = {https://www.aeaweb.org/articles?id=10.1257/app.1.3.1}}

@misc{hindenburg_research_opera_2020,
	title = {Opera: {Phantom} of the {Turnaround} – 70\% {Downside}},
	shorttitle = {Opera},
	url = {https://hindenburgresearch.com/opera-phantom-of-the-turnaround/},
	language = {en-US},
	urldate = {2021-08-13},
	journal = {Hindenburg Research},
	author = {{Hindenburg Research}},
	month = jan,
	year = {2020},
}

@article {klk,
	title = {Experimental Analysis of Neighborhood Effects},
	journal = {Econometrica},
	volume = {75},
	number = {1},
	year = {2007},
	pages = {83-119},
	author = {Jeffrey R. Kling and Jeffrey B. Liebman and Lawrence F. Katz}
}

@article{karlan_zinman_2010,
    author = {Karlan, Dean and Zinman, Jonathan},
    title = "{Expanding Credit Access: Using Randomized Supply Decisions to Estimate the Impacts}",
    journal = {The Review of Financial Studies},
    volume = {23},
    number = {1},
    pages = {433-464},
    year = {2010},
    month = {11},
    issn = {0893-9454},
    doi = {10.1093/rfs/hhp092},
    url = {https://doi.org/10.1093/rfs/hhp092},
    eprint = {https://academic.oup.com/rfs/article-pdf/23/1/433/24436375/hhp092.pdf},
}

@techreport{field19,
 title = "On Her Own Account: How Strengthening Women's Financial Control Affects Labor Supply and Gender Norms",
 author = "Field, Erica M and Pande, Rohini and Rigol, Natalia and Schaner, Simone G and Moore, Charity Troyer",
 institution = "National Bureau of Economic Research",
 type = "Working Paper",
 series = "Working Paper Series",
 number = "26294",
 year = "2019",
 month = "September",
 doi = {10.3386/w26294},
 URL = "http://www.nber.org/papers/w26294",
}

@techreport{cfpb,
	title = {{CFPB Financial Well-Being Scale}},
	url = {https://files.consumerfinance.gov/f/documents/201705_cfpb_financial-well-being-scale-technical-report.pdf},
	institution = {Consumer Financial Protection Bureau},
	author = {{Consumer Finance Protection Bureau}},
	date = {2017-05},
	year={2017}
}

@article{Ridley,
	author = {Ridley, Matthew and Rao, Gautam and Schilbach, Frank and Patel, Vikram},
	title = {Poverty, depression, and anxiety: Causal evidence and mechanisms},
	volume = {370},
	number = {6522},
	elocation-id = {eaay0214},
	year = {2020},
	doi = {10.1126/science.aay0214},
	publisher = {American Association for the Advancement of Science},
	issn = {0036-8075},
	URL = {https://science.sciencemag.org/content/370/6522/eaay0214},
	eprint = {https://science.sciencemag.org/content/370/6522/eaay0214.full.pdf},
	journal = {Science}
}

@article{anderson08,
 ISSN = {01621459},
 URL = {http://www.jstor.org/stable/27640197},
 author = {Michael L. Anderson},
 journal = {Journal of the American Statistical Association},
 number = {484},
 pages = {1481--1495},
 publisher = {[American Statistical Association, Taylor & Francis, Ltd.]},
 title = {Multiple Inference and Gender Differences in the Effects of Early Intervention: A Reevaluation of the Abecedarian, Perry Preschool, and Early Training Projects},
 volume = {103},
 year = {2008}
}

@techreport{malawi_kutchova,
	title = {Digital Credit: Filling a hole, or digging a hole? Evidence from Malawi},
	author = {Brailovskaya, Valentina and Dupas, Pascaline and Robinson, Jonathan},
	year = {2020},
	type={Working Paper}
}

@techreport{francis_digital_2017,
	title = {Digital Credit: A Snapshot of the Current Landscape and Open Research Questions},
	url = {http://ibread.org/bread/working/516},
	type={Working Paper}, 
	institution = {BREAD},
	number={516},
	author = {Francis, Eilin and Blumenstock, Joshua E. and Robinson, Jonathan},
	urldate = {2018-01-03},
	date = {2017-07},
	year = {2017}
}

@misc{boston_review,
	type = {Text},
	title = {Perpetual {Debt} in the {Silicon} {Savannah}},
	url = {https://bostonreview.net/class-inequality-global-justice/kevin-p-donovan-emma-park-perpetual-debt-silicon-savannah},
	language = {en},
	urldate = {2021-08-04},
	journal = {Boston Review},
	author = {Donovan, Kevin P. and Park, Emma},
	month = aug,
	year = {2019},
}

@article{bharadwaj_fintech_2019,
	title = {Fintech and {Household} {Resilience} to {Shocks}: {Evidence} from {Digital} {Loans} in {Kenya}},
	shorttitle = {Fintech and {Household} {Resilience} to {Shocks}},
	url = {https://www.nber.org/papers/w25604.ack},
	urldate = {2019-04-07},
	journal = {National Bureau of Economic Research Working Paper Series},
	author = {Bharadwaj, Prashant and Jack, William and Suri, Tavneet},
	month = feb,
	year = {2019},
}

@article{skiba_payday_2019,
	title = {Do Payday Loans Cause Bankruptcy?},
	volume = {62},
	number = {3},
	journal = {The Journal of Law and Economics},
	author = {Skiba, Paige Marta and Tobacman, Jeremy},
	year = {2019},
	month={august},
	doi={https://www.journals.uchicago.edu/doi/full/10.1086/706201},
	pages = {485--519},
}

@article{karlan_microcredit_2011,
	title = {Microcredit in {Theory} and {Practice}: {Using} {Randomized} {Credit} {Scoring} for {Impact} {Evaluation}},
	volume = {332},
	copyright = {Copyright © 2011, American Association for the Advancement of Science},
	issn = {0036-8075, 1095-9203},
	shorttitle = {Microcredit in {Theory} and {Practice}},
	url = {http://science.sciencemag.org/content/332/6035/1278},
	doi = {10.1126/science.1200138},
	language = {en},
	number = {6035},
	urldate = {2019-01-09},
	journal = {Science},
	author = {Karlan, Dean and Zinman, Jonathan},
	month = jun,
	year = {2011},
	pmid = {21659596},
	pages = {1278--1284},
}

@article{tarozzi_impacts_2015,
	title = {The {Impacts} of {Microcredit}: {Evidence} from {Ethiopia}},
	volume = {7},
	issn = {1945-7782},
	shorttitle = {The {Impacts} of {Microcredit}},
	url = {https://www.aeaweb.org/articles?id=10.1257/app.20130475},
	doi = {10.1257/app.20130475},
	language = {en},
	number = {1},
	urldate = {2019-01-09},
	journal = {American Economic Journal: Applied Economics},
	author = {Tarozzi, Alessandro and Desai, Jaikishan and Johnson, Kristin},
	month = jan,
	year = {2015},
	pages = {54--89},
}

@article{augsburg_impacts_2015,
	title = {The {Impacts} of {Microcredit}: {Evidence} from {Bosnia} and {Herzegovina}},
	volume = {7},
	issn = {1945-7782},
	shorttitle = {The {Impacts} of {Microcredit}},
	url = {https://www.aeaweb.org/articles?id=10.1257/app.20130272},
	doi = {10.1257/app.20130272},
	language = {en},
	number = {1},
	urldate = {2019-01-09},
	journal = {American Economic Journal: Applied Economics},
	author = {Augsburg, Britta and De Haas, Ralph and Harmgart, Heike and Meghir, Costas},
	month = jan,
	year = {2015},
	pages = {183--203},
}

@article{crepon_estimating_2015,
	title = {Estimating the {Impact} of {Microcredit} on {Those} {Who} {Take} {It} {Up}: {Evidence} from a {Randomized} {Experiment} in {Morocco}},
	volume = {7},
	issn = {1945-7782},
	shorttitle = {Estimating the {Impact} of {Microcredit} on {Those} {Who} {Take} {It} {Up}},
	url = {https://www.aeaweb.org/articles?id=10.1257/app.20130535},
	doi = {10.1257/app.20130535},
	language = {en},
	number = {1},
	urldate = {2019-01-09},
	journal = {American Economic Journal: Applied Economics},
	author = {Crépon, Bruno and Devoto, Florencia and Duflo, Esther and Parienté, William},
	month = jan,
	year = {2015},
	pages = {123--150},
}

@article{attanasio_impacts_2015,
	title = {The {Impacts} of {Microfinance}: {Evidence} from {Joint}-{Liability} {Lending} in {Mongolia}},
	volume = {7},
	issn = {1945-7782},
	shorttitle = {The {Impacts} of {Microfinance}},
	url = {https://www.aeaweb.org/articles?id=10.1257/app.20130489},
	doi = {10.1257/app.20130489},
	language = {en},
	number = {1},
	urldate = {2019-01-09},
	journal = {American Economic Journal: Applied Economics},
	author = {Attanasio, Orazio and Augsburg, Britta and De Haas, Ralph and Fitzsimons, Emla and Harmgart, Heike},
	month = jan,
	year = {2015},
	pages = {90--122},
}

@article{banerjee_miracle_2015,
	title = {The {Miracle} of {Microfinance}? {Evidence} from a {Randomized} {Evaluation}},
	volume = {7},
	issn = {1945-7782},
	shorttitle = {The {Miracle} of {Microfinance}?},
	url = {https://www.aeaweb.org/articles?id=10.1257/app.20130533},
	doi = {10.1257/app.20130533},
	language = {en},
	number = {1},
	urldate = {2019-01-09},
	journal = {American Economic Journal: Applied Economics},
	author = {Banerjee, Abhijit and Duflo, Esther and Glennerster, Rachel and Kinnan, Cynthia},
	month = jan,
	year = {2015},
	pages = {22--53},
}

@article{angelucci_microcredit_2015,
	title = {Microcredit {Impacts}: {Evidence} from a {Randomized} {Microcredit} {Program} {Placement} {Experiment} by {Compartamos} {Banco}},
	volume = {7},
	issn = {1945-7782},
	shorttitle = {Microcredit {Impacts}},
	url = {https://www.aeaweb.org/articles?id=10.1257/app.20130537},
	doi = {10.1257/app.20130537},
	language = {en},
	number = {1},
	urldate = {2019-01-09},
	journal = {American Economic Journal: Applied Economics},
	author = {Angelucci, Manuela and Karlan, Dean and Zinman, Jonathan},
	month = jan,
	year = {2015},
	pages = {151--182},
}

@article{meager_understanding_2019,
	title = {Understanding the {Average} {Impact} of {Microcredit} {Expansions}: {A} {Bayesian} {Hierarchical} {Analysis} of {Seven} {Randomized} {Experiments}},
	volume = {11},
	shorttitle = {Understanding the {Average} {Impact} of {Microcredit} {Expansions}},
	url = {https://ideas.repec.org/a/aea/aejapp/v11y2019i1p57-91.html},
	language = {en},
	number = {1},
	urldate = {2019-01-09},
	journal = {American Economic Journal: Applied Economics},
	author = {Meager, Rachael},
	year = {2019},
	pages = {57--91},
}

@techreport{allcott_are_2021,
	type = {Working {Paper}},
	title = {Are {High}-{Interest} {Loans} {Predatory}? {Theory} and {Evidence} from {Payday} {Lending}},
	shorttitle = {Are {High}-{Interest} {Loans} {Predatory}?},
	url = {https://www.nber.org/papers/w28799},
	number = {28799},
	urldate = {2021-08-03},
	institution = {National Bureau of Economic Research},
	author = {Allcott, Hunt and Kim, Joshua J. and Taubinsky, Dmitry and Zinman, Jonathan},
	month = may,
	year = {2021},
	doi = {10.3386/w28799},
	note = {Series: Working Paper Series},
}

@article{bhutta_consumer_2016,
	title = {Consumer {Borrowing} after {Payday} {Loan} {Bans}},
	volume = {59},
	issn = {0022-2186},
	url = {https://www.journals.uchicago.edu/doi/10.1086/686033},
	doi = {10.1086/686033},
	number = {1},
	urldate = {2021-08-03},
	journal = {The Journal of Law and Economics},
	author = {Bhutta, Neil and Goldin, Jacob and Homonoff, Tatiana},
	month = feb,
	year = {2016},
	note = {Publisher: The University of Chicago Press},
	pages = {225--259},
}

@article{bhutta_payday_2015,
	title = {Payday {Loan} {Choices} and {Consequences}},
	volume = {47},
	issn = {1538-4616},
	url = {https://onlinelibrary.wiley.com/doi/abs/10.1111/jmcb.12175},
	doi = {10.1111/jmcb.12175},
	language = {en},
	number = {2-3},
	urldate = {2021-08-03},
	journal = {Journal of Money, Credit and Banking},
	author = {Bhutta, Neil and Skiba, Paige Marta and Tobacman, Jeremy},
	year = {2015},
	note = {\_eprint: https://onlinelibrary.wiley.com/doi/pdf/10.1111/jmcb.12175},
	pages = {223--260},
}

@article{melzer_real_2011,
	title = {The {Real} {Costs} of {Credit} {Access}: {Evidence} from the {Payday} {Lending} {Market}*},
	volume = {126},
	issn = {0033-5533},
	shorttitle = {The {Real} {Costs} of {Credit} {Access}},
	url = {https://doi.org/10.1093/qje/qjq009},
	doi = {10.1093/qje/qjq009},
	number = {1},
	urldate = {2021-08-03},
	journal = {The Quarterly Journal of Economics},
	author = {Melzer, Brian T.},
	month = feb,
	year = {2011},
	pages = {517--555},
}

@article{melzer_spillovers_2018,
	title = {Spillovers from {Costly} {Credit}},
	volume = {31},
	issn = {0893-9454},
	url = {https://doi.org/10.1093/rfs/hhx134},
	doi = {10.1093/rfs/hhx134},
	number = {9},
	urldate = {2021-08-03},
	journal = {The Review of Financial Studies},
	author = {Melzer, Brian T},
	month = sep,
	year = {2018},
	pages = {3568--3594},
}

@article{morgan_how_2012,
	title = {How {Payday} {Credit} {Access} {Affects} {Overdrafts} and {Other} {Outcomes}},
	volume = {44},
	issn = {0022-2879},
	url = {https://www.jstor.org/stable/41487807},
	number = {2/3},
	urldate = {2021-08-03},
	journal = {Journal of Money, Credit and Banking},
	author = {Morgan, Donald P. and Strain, Michael R. and Seblani, Ihab},
	year = {2012},
	note = {Publisher: Wiley},
	pages = {519--531},
}

@article{morse_payday_2011,
	title = {Payday lenders: {Heroes} or villains?},
	volume = {102},
	issn = {0304-405X},
	shorttitle = {Payday lenders},
	url = {https://www.sciencedirect.com/science/article/pii/S0304405X11000870},
	doi = {10.1016/j.jfineco.2011.03.022},
	language = {en},
	number = {1},
	urldate = {2021-08-03},
	journal = {Journal of Financial Economics},
	author = {Morse, Adair},
	month = oct,
	year = {2011},
	pages = {28--44},
}

@article{zinman_restricting_2010,
	title = {Restricting consumer credit access: {Household} survey evidence on effects around the {Oregon} rate cap},
	volume = {34},
	shorttitle = {Restricting consumer credit access},
	url = {https://ideas.repec.org/a/eee/jbfina/v34y2010i3p546-556.html},
	language = {en},
	number = {3},
	urldate = {2021-08-03},
	journal = {Journal of Banking \& Finance},
	author = {Zinman, Jonathan},
	year = {2010},
	note = {Publisher: Elsevier},
	pages = {546--556},
}

@article{carrell_harms_2014,
	title = {In {Harm}'s {Way}? {Payday} {Loan} {Access} and {Military} {Personnel} {Performance}},
	volume = {27},
	issn = {0893-9454},
	shorttitle = {In {Harm}'s {Way}?},
	url = {https://doi.org/10.1093/rfs/hhu034},
	doi = {10.1093/rfs/hhu034},
	number = {9},
	urldate = {2021-08-03},
	journal = {The Review of Financial Studies},
	author = {Carrell, Scott and Zinman, Jonathan},
	month = sep,
	year = {2014},
	pages = {2805--2840},
}

@article{gathergood_how_2019,
	title = {How {Do} {Payday} {Loans} {Affect} {Borrowers}? {Evidence} from the {U}.{K}. {Market}},
	volume = {32},
	issn = {0893-9454},
	shorttitle = {How {Do} {Payday} {Loans} {Affect} {Borrowers}?},
	url = {https://doi.org/10.1093/rfs/hhy090},
	doi = {10.1093/rfs/hhy090},
	number = {2},
	urldate = {2021-08-03},
	journal = {The Review of Financial Studies},
	author = {Gathergood, John and Guttman-Kenney, Benedict and Hunt, Stefan},
	month = feb,
	year = {2019},
	pages = {496--523},
}

@article{johnen_promises_2021,
	title = {Promises and pitfalls of digital credit: {Empirical} evidence from {Kenya}},
	volume = {16},
	issn = {1932-6203},
	shorttitle = {Promises and pitfalls of digital credit},
	url = {https://journals.plos.org/plosone/article?id=10.1371/journal.pone.0255215},
	doi = {10.1371/journal.pone.0255215},
	language = {en},
	number = {7},
	urldate = {2021-08-04},
	journal = {PLOS ONE},
	author = {Johnen, Constantin and Parlasca, Martin and Mußhoff, Oliver},
	month = jul,
	year = {2021},
	note = {Publisher: Public Library of Science},
	pages = {e0255215},
}

@article{bjorkegren_behavior_2020,
	title = {Behavior {Revealed} in {Mobile} {Phone} {Usage} {Predicts} {Credit} {Repayment}},
	volume = {34},
	issn = {0258-6770},
	url = {https://doi.org/10.1093/wber/lhz006},
	doi = {10.1093/wber/lhz006},
	number = {3},
	urldate = {2021-08-04},
	journal = {The World Bank Economic Review},
	author = {Björkegren, Daniel and Grissen, Darrell},
	month = oct,
	year = {2020},
	pages = {618--634},
}

@misc{totolo_kenyas_2018,
	type = {{CGAP}},
	title = {Kenya’s {Digital} {Credit} {Revolution} {Five} {Years} {On}},
	url = {https://www.cgap.org/blog/kenyas-digital-credit-revolution-five-years},
	language = {en},
	urldate = {2021-08-04},
	journal = {Kenya’s Digital Credit Revolution Five Years On},
	author = {Totolo, Edoardo},
	month = apr,
	year = {2018},
}
